\documentclass[11pt, a4paper]{article}
\usepackage{jheppub, slashed, color,subfig}
\usepackage{enumitem}
\usepackage{multirow}
\usepackage{tikz}
\usepackage{amsmath}
\allowdisplaybreaks[2]
\usepackage{color}
\usepackage[multiple]{footmisc}%for 

\usepackage{textcomp}
\usepackage{amssymb}%for semidirect product symbol
\usetikzlibrary{positioning}
\usetikzlibrary{calc,intersections,through,backgrounds}

\usepackage{tikz-3dplot}
\usepackage{amscd}
\usepackage{amsthm}
\usepackage{array} 
\usepackage{graphicx}
\usepackage{adjustbox}
\usepackage{mathrsfs}
\usepackage{dsfont}
\newcommand{\C}{\mathbb{C}}
\usepackage{tikz}
\usetikzlibrary{decorations.pathmorphing}

\DeclareFontFamily{U}{MnSymbolC}{}
\DeclareSymbolFont{MnSyC}{U}{MnSymbolC}{m}{n}
\DeclareFontShape{U}{MnSymbolC}{m}{n}{
    <-6>  MnSymbolC5
   <6-7>  MnSymbolC6
   <7-8>  MnSymbolC7
   <8-9>  MnSymbolC8
   <9-10> MnSymbolC9
  <10-12> MnSymbolC10
  <12->   MnSymbolC12}{}
\DeclareMathSymbol{\intprod}{\mathbin}{MnSyC}{'270}

\newcommand{\til}{\widetilde}

\newcommand{\zb}{\bar{z}}

\let\nc\newcommand
\let\renc\renewcommand
\nc{\wbar}{\overline}
\let\td\tilde
\let\wtd\widetilde
\let\wht\widehat
\let\mcl\mathcal

\nc{\ab}{{\bar{a}}} \nc{\at}{\tilde{a}} \nc{\ah}{\hat{a}}
\nc{\bb}{{\bar{b}}} 
\nc{\cb}{{\bar{c}}} \nc{\ct}{\tilde{c}} %\nc{\ch}{\hat{c}}
\nc{\fb}{{\bar{f}}} \nc{\ft}{\tilde{f}} \nc{\fh}{\hat{f}}
\nc{\gb}{{\bar{g}}} \nc{\gt}{\tilde{g}} \nc{\gh}{\hat{g}}
\nc{\hb}{{\bar{h}}} \nc{\hh}{\hat{h}} %\nc{\ht}{\tilde{h}}
\nc{\ib}{{\bar{\imath}}} \nc{\ih}{\hat{\imath}} %\nc{\it}{\tilde{\imath}}
\nc{\jb}{{\bar{\jmath}}} \nc{\jt}{\tilde{\jmath}} \nc{\jh}{\hat{\jmath}}
\nc{\kb}{{\bar{k}}} \nc{\kt}{\tilde{k}} \nc{\kh}{\hat{k}}
\nc{\lb}{{\bar{l}}} \nc{\lt}{\tilde{l}} \nc{\lh}{\hat{l}}
\nc{\mb}{{\bar{m}}} \nc{\mt}{\tilde{m}} \nc{\mh}{\hat{m}}
\nc{\nb}{{\bar{n}}} \nc{\nt}{\tilde{n}} \nc{\nh}{\hat{n}}
\nc{\ob}{{\bar{o}}} \nc{\ot}{\tilde{o}} \nc{\oh}{\hat{o}}
\nc{\pb}{{\bar{p}}} \nc{\pt}{\tilde{p}} \nc{\ph}{\hat{p}}
\nc{\qb}{{\bar{q}}} \nc{\qt}{\tilde{q}} \nc{\qh}{\hat{q}}
\nc{\rb}{{\bar{r}}} \nc{\rt}{\tilde{r}} 
\renc{\sb}{{\bar{s}}} \nc{\st}{\tilde{s}} \nc{\sh}{\hat{s}}
\nc{\tb}{{\bar{t}}} \renc{\th}{\hat{t}} %\nc{\tt}{\tilde{t}}
\nc{\ub}{{\bar{u}}} \nc{\ut}{\tilde{u}} \nc{\uh}{\hat{u}}
\nc{\vb}{{\bar{v}}} \nc{\vt}{\tilde{v}} \nc{\vh}{\hat{v}}
\nc{\wb}{{\bar{w}}} \nc{\wt}{\tilde{w}} \nc{\wh}{\hat{w}}
\nc{\xb}{{\bar{x}}} \nc{\xt}{\tilde{x}} \nc{\xh}{\hat{x}}
\nc{\yb}{{\bar{y}}} \nc{\yt}{\tilde{y}} \nc{\yh}{\hat{y}}
\nc{\Ab}{\wbar{A}} \nc{\At}{\wtd{A}} \nc{\Ah}{\wht{A}}
\nc{\Bb}{\wbar{B}} \nc{\Bt}{\wtd{B}} \nc{\Bh}{\wht{B}}
\nc{\Cb}{\wbar{C}} \nc{\Ct}{\wtd{C}} \nc{\Ch}{\wht{C}}
\nc{\Db}{\wbar{D}} \nc{\Dt}{\wtd{D}} \nc{\Dh}{\wht{D}}
\nc{\Eb}{\wbar{E}} \nc{\Et}{\wtd{E}} \nc{\Eh}{\wht{E}}
\nc{\Fb}{\wbar{F}} \nc{\Ft}{\wtd{F}} \nc{\Fh}{\wht{F}}
\nc{\Gb}{\wbar{G}} \nc{\Gt}{\wtd{G}} \nc{\Gh}{\wht{G}}
\nc{\Hb}{\wbar{H}} \nc{\Ht}{\wtd{H}} \nc{\Hh}{\wht{H}}
\nc{\Ib}{\wbar{I}} \nc{\It}{\wtd{I}} \nc{\Ih}{\wht{I}}
\nc{\Jb}{\wbar{J}} \nc{\Jt}{\wtd{J}} \nc{\Jh}{\wht{J}}
\nc{\Kb}{\wbar{K}} \nc{\Kt}{\wtd{K}} \nc{\Kh}{\wht{K}}
\nc{\Lb}{\wbar{L}} \nc{\Lt}{\wtd{L}} \nc{\Lh}{\wht{L}}
\nc{\Mb}{\wbar{M}} \nc{\Mt}{\wtd{M}} \nc{\Mh}{\wht{M}}
\nc{\Nb}{\wbar{N}} \nc{\Nt}{\wtd{N}} \nc{\Nh}{\wht{N}}
\nc{\Ob}{\wbar{O}} \nc{\Ot}{\wtd{O}} \nc{\Oh}{\wht{O}}
\nc{\Pb}{\wbar{P}} \nc{\Pt}{\wtd{P}} \nc{\Ph}{\wht{P}}
\nc{\Qb}{\wbar{Q}} \nc{\Qt}{\wtd{Q}} \nc{\Qh}{\wht{Q}}
\nc{\Rb}{\wbar{R}} \nc{\Rt}{\wtd{R}} \nc{\Rh}{\wht{R}}
\nc{\Sb}{\wbar{S}} \nc{\St}{\wtd{S}} \nc{\Sh}{\wht{S}}
\nc{\Tb}{\wbar{T}} \nc{\Tt}{\wtd{T}} \nc{\Th}{\wht{T}}
\nc{\Ub}{\wbar{U}} \nc{\Ut}{\wtd{U}} \nc{\Uh}{\wht{U}}
\nc{\Vb}{\wbar{V}} \nc{\Vt}{\wtd{V}} \nc{\Vh}{\wht{V}}
\nc{\Wb}{\wbar{W}} \nc{\Wt}{\wtd{W}} \nc{\Wh}{\wht{W}}
\nc{\Xb}{\wbar{X}} \nc{\Xt}{\wtd{X}} \nc{\Xh}{\wht{X}}
\nc{\Yb}{\wbar{Y}} \nc{\Yt}{\wtd{Y}} \nc{\Yh}{\wht{Y}}
\nc{\Zb}{\wbar{Z}} \nc{\Zt}{\wtd{Z}} \nc{\Zh}{\wht{Z}}

\nc{\CA}{\mcl{A}} \nc{\CAb}{\wbar{\CA}} \nc{\CAt}{\wtd{\CA}} \nc{\CAh}{\wht{\CA}}
\nc{\CB}{\mcl{B}} \nc{\CBb}{\wbar{\CB}} \nc{\CBt}{\wtd{\CB}} \nc{\CBh}{\wht{\CB}}
\nc{\cD}{\mcl{D}} \nc{\cDb}{\wbar{\cD}} \nc{\cDt}{\wtd{\cC}} \nc{\cDh}{\wht{\cD}}
\nc{\CE}{\mcl{E}} \nc{\CEb}{\wbar{\CE}} \nc{\CEt}{\wtd{\CE}} \nc{\CEh}{\wht{\CE}}
\nc{\CF}{\mcl{F}} \nc{\CFb}{\wbar{\CF}} \nc{\CFt}{\wtd{\CF}} \nc{\CFh}{\wht{\CF}}
\nc{\CG}{\mcl{G}} \nc{\CGb}{\wbar{\CG}} \nc{\CGt}{\wtd{\CG}} \nc{\CGh}{\wht{\CG}}
\nc{\CH}{\mcl{H}} \nc{\CHb}{\wbar{\CH}} \nc{\CHt}{\wtd{\CH}} \nc{\CHh}{\wht{\CH}}
\nc{\CI}{\mcl{I}} \nc{\CIb}{\wbar{\CI}} \nc{\CIt}{\wtd{\CI}} \nc{\CIh}{\wht{\CI}}
\nc{\CJ}{\mcl{J}} \nc{\CJb}{\wbar{\CJ}} \nc{\CJt}{\wtd{\CJ}} \nc{\CJh}{\wht{\CJ}}
\nc{\CK}{\mcl{K}} \nc{\CKb}{\wbar{\CK}} \nc{\CKt}{\wtd{\CK}} \nc{\CKh}{\wht{\CK}}
\nc{\CL}{\mcl{L}} \nc{\CLb}{\wbar{\CL}} \nc{\CLt}{\wtd{\CL}} \nc{\CLh}{\wht{\CL}}
\nc{\CM}{\mcl{M}} \nc{\CMb}{\wbar{\CM}} \nc{\CMt}{\wtd{\CM}} \nc{\CMh}{\wht{\CM}}
\nc{\CN}{\mcl{N}} \nc{\CNb}{\wbar{\CN}} \nc{\CNt}{\wtd{\CN}} \nc{\CNh}{\wht{\CN}}
\nc{\CO}{\mcl{O}} \nc{\COb}{\wbar{\CO}} \nc{\COt}{\wtd{\CO}} \nc{\COh}{\wht{\CO}}
\nc{\CQ}{\mcl{Q}} \nc{\CQb}{\wbar{\CQ}} \nc{\CQt}{\wtd{\CQ}} \nc{\CQh}{\wht{\CQ}}
\nc{\CR}{\mcl{R}} \nc{\CRb}{\wbar{\CR}} \nc{\CRt}{\wtd{\CR}} \nc{\CRh}{\wht{\CR}}
\nc{\CS}{\mcl{S}} \nc{\CSb}{\wbar{\CS}} \nc{\CSt}{\wtd{\CS}} \nc{\CSh}{\wht{\CS}}
\nc{\CT}{\mcl{T}} \nc{\CTb}{\wbar{\CT}} \nc{\CTt}{\wtd{\CT}} \nc{\CTh}{\wht{\CT}}
\nc{\CU}{\mcl{U}} \nc{\CUb}{\wbar{\CU}} \nc{\CUt}{\wtd{\CU}} \nc{\CUh}{\wht{\CU}}
\nc{\CV}{\mcl{V}} \nc{\CVb}{\wbar{\CV}} \nc{\CVt}{\wtd{\CV}} \nc{\CVh}{\wht{\CV}}
\nc{\CW}{\mcl{W}} \nc{\CWb}{\wbar{\CW}} \nc{\CWt}{\wtd{\CW}} \nc{\CWh}{\wht{\CW}}
\nc{\CX}{\mcl{X}} \nc{\CXb}{\wbar{\CX}} \nc{\CXt}{\wtd{\CX}} \nc{\CXh}{\wht{\CX}}
\nc{\CY}{\mcl{Y}} \nc{\CYb}{\wbar{\CY}} \nc{\CYt}{\wtd{\CY}} \nc{\CYh}{\wht{\CY}}
\nc{\CZ}{\mcl{Z}} \nc{\CZb}{\wbar{\CZ}} \nc{\CZt}{\wtd{\CZ}} \nc{\CZh}{\wht{\CZ}}

\let\eps\epsilon
\let\ups\upsilon

\let\veps\varepsilon
\let\vtht\vartheta
\let\vsgm\varsigma
\let\vphi\varphi
\let\vrho\varrho

\nc{\alphab}{\bar{\alpha}} \nc{\alphat}{\td{\alpha}} \nc{\alphah}{\hat{\alpha}}
\nc{\betab}{\bar{\beta}}   \nc{\betat}{\td{\beta}}   \nc{\betah}{\hat{\beta}} 
\nc{\gammab}{\bar{\gamma}} \nc{\gammat}{\td{\gamma}} \nc{\gammah}{\hat{\gamma}} 
\nc{\deltab}{\bar{\delta}} \nc{\deltat}{\td{\delta}} \nc{\deltah}{\hat{\delta}} 
\nc{\epsilonb}{\bar{\eps}} \nc{\epsilont}{\td{\eps}} \nc{\epsilonh}{\hat{\eps}} 
\nc{\vepsb}{\bar{\veps}}   \nc{\vepst}{\td{\veps}}   \nc{\vepsh}{\hat{\veps}} 
\nc{\zetab}{\bar{\zeta}}   \nc{\zetat}{\td{\zeta}}   \nc{\zetah}{\hat{\zeta}} 
\nc{\etab}{\bar{\eta}}     \nc{\etat}{\td{\eta}}     \nc{\etah}{\hat{\eta}} 
\nc{\thetab}{\bar{\theta}} \nc{\thetat}{\td{\theta}} \nc{\thetah}{\hat{\theta}} 
\nc{\vthetab}{\bar{\vtht}} \nc{\vthetat}{\td{\vtht}} \nc{\vthetah}{\hat{\vtht}} 
\nc{\lambdab}{\bar{\lambda}} \nc{\lambdat}{\td{\lambda}} \nc{\lambdah}{\hat{\lambda}} 
\nc{\iotab}{\bar{\iota}}   \nc{\iotat}{\td{\iota}}   \nc{\iotah}{\hat{\iota}} 
\nc{\kappab}{\bar{\kappa}} \nc{\kappat}{\td{\kappa}} \nc{\kappah}{\hat{\kappa}} 
\nc{\lmdb}{\bar{\lmd}}     \nc{\lmdt}{\td{\lmd}}     \nc{\lmdh}{\hat{\lmd}} 
\nc{\mub}{\bar{\mu}}       \nc{\mut}{\td{\mu}}       \nc{\muh}{\hat{\mu}} 
\nc{\nub}{\bar{\nu}}       \nc{\nut}{\td{\nu}}       \nc{\nuh}{\hat{\nu}} 
\nc{\xib}{\bar{\xi}}       \nc{\xit}{\td{\xi}}       \nc{\xih}{\hat{\xi}} 
\nc{\pib}{\bar{\pi}}       \nc{\pit}{\td{\pi}}       \nc{\pih}{\hat{\pi}} 
\nc{\vpib}{\bar{\vpi}}     \nc{\vpit}{\td{\vpi}}     \nc{\vpih}{\hat{\vpi}} 
\nc{\rhob}{\bar{\rho}}     \nc{\rhot}{\td{\rho}}     \nc{\rhoh}{\hat{\rho}} 
\nc{\vrhob}{\bar{\vrho}}   \nc{\vrhot}{\td{\vrho}}   \nc{\vrhoh}{\hat{\vrho}} 
\nc{\sigmab}{\bar{\sigma}} \nc{\sigmat}{\td{\sigma}} \nc{\sigmah}{\hat{\sigma}} 
\nc{\vsigmab}{\bar{\vsgm}} \nc{\vsigmat}{\td{\vsgm}} \nc{\vsigmah}{\hat{\vsgm}} 
\nc{\taub}{\bar{\tau}}     \nc{\taut}{\td{\tau}}     \nc{\tauh}{\hat{\tau}} 
\nc{\upsb}{\bar{\ups}} \nc{\upst}{\td{\ups}} \nc{\upsh}{\hat{\ups}} 
\nc{\phib}{\bar{\phi}}     \nc{\phit}{\td{\phi}}     \nc{\phih}{\hat{\phi}} 
\nc{\varphib}{\bar{\vphi}}   \nc{\varphit}{\td{\vphi}}   \nc{\varphih}{\hat{\vphi}} 
\nc{\chib}{\bar{\chi}}     \nc{\chit}{\td{\chi}}     \nc{\chih}{\hat{\chi}} 
\nc{\omegab}{\bar{\omega}} \nc{\omegat}{\td{\omega}} \nc{\omegah}{\hat{\omega}} 

\nc{\Gammab}{\wbar{\Gamma}}     \nc{\Gammat}{\wtd{\Gamma}}     \nc{\Gammah}{\wht{\Gamma}}
\nc{\Deltab}{\wbar{\Delta}}     \nc{\Deltat}{\wtd{\Delta}}     \nc{\Deltah}{\wht{\Delta}}
\nc{\Thetab}{\wbar{\Theta}}     \nc{\Thetat}{\wtd{\Theta}}     \nc{\Thetah}{\wht{\Theta}}
\nc{\Lambdab}{\wbar{\Lambda}}   \nc{\Lambdat}{\wtd{\Lambda}}   \nc{\Lambdah}{\wht{\Lambda}}
\nc{\Xib}{\wbar{\Xi}}           \nc{\Xit}{\wtd{\Xi}}           \nc{\Xih}{\wht{\Xi}}
\nc{\Pib}{\wbar{\Pi}}           \nc{\Pit}{\wtd{\Pi}}           \nc{\Pih}{\wht{\Pi}}
\nc{\Sigmab}{\wbar{\Sigma}}     \nc{\Sigmat}{\wtd{\Sigma}}     \nc{\Sigmah}{\wht{\Sigma}}
\nc{\Upsilonb}{\wbar{\Upsilon}} \nc{\Upsilont}{\wtd{\Upsilon}} \nc{\Upsilonh}{\wht{\Upsilon}}
\nc{\Phib}{\wbar{\Phi}}         \nc{\Phit}{\wtd{\Phi}}         \nc{\Phih}{\wht{\Phi}}
\nc{\Psib}{\wbar{\Psi}}         \nc{\Psit}{\wtd{\Psi}}         \nc{\Psih}{\wht{\Psi}}
\nc{\Omegab}{\wbar{\Omega}}     \nc{\Omegat}{\wtd{\Omega}}     \nc{\Omegah}{\wht{\Omega}}

\nc{\txd}{d}

\nc{\tcos}{\textrm{cos}}
\nc{\tsin}{\textrm{sin}}

\nc{\al}{\alpha}
\nc{\bt}{\beta}
\nc{\gm}{\gamma}
\nc{\rh}{\rho}
\nc{\zt}{\zeta}
\nc{\Dl}{\Delta}
\nc{\TD}{\til{\Delta}}

\nc{\sg}{\Sigma}

\nc{\rd}{0.75}

\newcommand{\pa}{\partial}
\newcommand{\D}{\delta}

\def\ie{\begin{equation}\begin{aligned}}
\def\fe{\end{aligned}\end{equation}}
\begin{document}

\title{Intersecting Surface Operators in 6d Holomorphic Field Theories }
\author{Meer Ashwinkumar}

\affiliation{Albert Einstein Center for Fundamental Physics, Institute for Theoretical Physics,
University of Bern, Sidlerstrasse 5, CH-3012 Bern, Switzerland}

\emailAdd{meer.ashwinkumar@unibe.ch}

\abstract{We study intersecting surface operators in 6d holomorphic field theories with the aim of unraveling associated quantum integrable structures. We first study the intersections of surface operators in 6d holomorphic Chern--Simons theory on $\C^3$. Computing their correlation function, 
we find a local operator at the intersection of the surface operators with a form reminiscent of the leading nontrivial term in the quasi-classical expansion of a rational $R$-matrix, as predicted by Costello.   
We provide evidence that this $R$-matrix-like operator satisfies a Yang--Baxter-type relation. We then derive the associated coproduct of the chiral algebra supported by surface operators from their OPE.
We also study intersecting surface operators in 6d holomorphic BF theory and derive the local leading form of the corresponding $R$-matrix-like
operator. When this theory is placed on twistor space, where it describes the
self-dual sector of 4d Yang--Mills theory, this operator is expected to provide
a local building block for quantum integrable structures anticipated in that setting.}

\maketitle

\noindent
\section{Introduction}

Six-dimensional holomorphic Chern--Simons (CS) theory is known to describe the open topological B-string ending on a space-filling brane \cite{Witten:1992fb}. It has the action 
\ie\label{act1}
\frac{i}{8\pi}\int_{\C^3} dv\wedge dw \wedge dz \wedge \textrm{Tr} \left(A\wedge dA +\frac{2}{3}A\wedge A \wedge A \right),
\fe
where $v,w,z$ are holomorphic coordinates on $\C^3$, and $A$ is the partial connection 
\begin{equation}
A=A_{\vb}d\vb  + A_{\wb} d\wb  +A_{\zb} d\zb.
\end{equation}
In recent years, there has been much interest in 6d CS from the perspective of integrability \cite{Bittleston:2020hfv,Penna:2020uky,Cole:2023umd} as well as connections to celestial holography \cite{Costello:2022wso, Costello:2023hmi}. The aim of this work is to understand quantum integrability in 6d CS as well as in a related theory, namely, 6d holomorphic BF theory. The latter theory is especially important, as it is equivalent to the self-dual sector of 4d Yang--Mills theory when defined on twistor space.

In this work, one of our main aims is to study correlation functions of  intersections of surface operators in 6d holomorphic Chern--Simons theory and 6d holomorphic BF theory, which are interpreted as B-branes in the former. We are motivated by a conjecture of Costello \cite{Costello:2021bah}, which says that the correlation function of intersecting surface operators in 6d CS on twistor space is similar in form to the quasi-classical expansion of a rational, spectral parameter-dependent $R$-matrix, but with matrices replaced by currents. 

In \cite{Costello:2021bah}, Costello briefly remarked that when two surface operators of 6d Chern--Simons theory on twistor space intersect, while located at points $z_1$ and $z_2$ on the twistor sphere, one expects correlation functions exhibiting a singularity of the schematic form 
\ie \label{jjopeint}
1+ \frac{\hbar}{z_1-z_2}J^a \otimes J'_a + O(\hbar^2),
\fe 
where $J$ and $J'$ are currents associated with each of the surface operators, respectively. 
However, no detailed computation of these correlators or their algebraic consequences was carried out in \cite{Costello:2021bah}. One of our main goals is to provide an explicit derivation of \eqref{jjopeint}, at least in some simplifying limits.

In particular, we shall confirm the formula \eqref{jjopeint} when twistor space is replaced by $\mathbb{C}^3$, and we moreover derive an analogous result for 6d holomorphic BF theory on $\mathbb{C}^3$. Although we work on $\mathbb{C}^3$ instead of twistor space, we expect our results to generalize when studying twistor space. Crucially, we expect that the $R$-matrix we derive for 6d holomorphic BF theory will generalize in a natural manner when considering twistor space, and such an $R$-matrix would underlie the quantum integrability of the self-dual sector of 4d Yang--Mills theory. 

The form of \eqref{jjopeint} is reminiscent of
a rational spectral parameter-dependent $R$-matrix that can be derived from intersecting line operators in 4d Chern--Simons theory \cite{Costello:2013zra,Costello:2017dso},
which has the action 
\begin{equation}\label{4dCS}
S=\frac{i}{2\pi\hbar}\int_{ \mathbb{R}^2\times \mathbb{C}} dz \wedge \textrm{Tr}\bigg(A\wedge d A + \frac{2}{3}  A\wedge  A\wedge  A\bigg).
\end{equation} 
In this theory, the correlation function of  line operators at points $z_1$ and $z_2$ on $\C$, and in representations $R_1$ and $R_2$ of the generators of $G$ gives rise to the $R$-matrix
\begin{equation}
\til{R}_{12}(z_1-z_2)=\mathds{1}+\frac{\hbar}{z_1-z_2}T_{R_1}^a\otimes T_{ R_2 a}+{O}(\hbar^2)
\end{equation}
to linear order in $\hbar$. 
The result \eqref{jjopeint} is thus natural given that the 4d, 5d and 6d Chern--Simons theories are related via a field-theoretic T-duality \cite{Yamazaki:2019prm},\footnote{A straightforward approach to understanding this relationship is by noticing that if we replace two of the complex planes in the action \eqref{act1} by cylinders $S^1 \times \mathbb{R}$, and dimensionally reduce along the circle directions, one ends up with the 4d Chern--Simons theory action \eqref{4dCS}. } and both 4d and 5d Chern--Simons theories realise $R$-matrices or generalizations thereof via intersecting operators \cite{Ashwinkumar:2024vys,Ishtiaque:2024orn}. 

However, a crucial difference from 4d Chern--Simons theory is that both 6d holomorphic Chern--Simons theory and 6d holomorphic BF theory suffer from gauge anomalies, that need to be cancelled via a Green--Schwarz mechanism \cite{Costello:2015xsa, Costello:2021bah, Costello:2019jsy, Costello:2018zrm, Costello:2023hmi}. Notably, this restricts the gauge group for 6d CS to be $G=SO(8)$ and the gauge group for 6d BF on twistor space to be $SU(2)$, $SU(3)$, $SO(8)$ or one of the exceptional groups. The Green--Schwarz mechanism involves closed string fields from the topological string perspective, which must be taken into account in the analysis. The action governing these closed string fields takes the form of BCOV/Kodaira-Spencer theory.
In this work, we shall mostly focus on the contributions of the open string sector to the intersection of surface operators, and discuss closed string effects sparingly. 

Let us provide a brief overview of this work. In Section 2, we review some known facts about 6d holomorphic Chern--Simons theory and its surface operators. In Section 3, we compute the leading order contribution to the correlation function of intersecting holomorphic surface operators, and provide evidence that the local operator arising at this intersection may be interpreted as an analogue of an $R$-matrix satisfying a Yang--Baxter equation. In addition, we compute the OPE of these surface operators/B-branes, which corresponds to a coproduct for the algebra of local operators on these operators. 
In Section 4, we derive the analogue of an  $R$-matrix from intersecting surface operators in 6d holomorphic BF theory, and further discuss implications for 6d BF theory on twistor space.

 \acknowledgments

    The author would like to thank Kevin Costello for helpful comments on a previous version of this manuscript. 
 	The author would also like to thank Roland Bittleston for an explanation on surface operators in 6d holomorphic BF theory, and  Matthias Blau for explanations on the tangent bundle group $TG$ and its associated Chern--Simons theory.   
This work is supported in part by the NCCR SwissMAP of the Swiss National Science Foundation.

\section{Review of 6d Holomorphic Chern--Simons Theory }

\subsection{Coupling to BCOV Theory and Holomorphic Surface Operators}\label{2.1}

In this subsection, we shall review some relevant aspects of holomorphic Chern--Simons theory. Firstly, as explained in \cite{Costello:2015xsa, Costello:2019jsy}, 6d holomorphic Chern--Simons theory defined by the action 
\ie 
\frac{i}{8\pi}\int_{\C^3} dv\wedge dw \wedge dz \wedge \textrm{Tr} \left(A\wedge dA +\frac{2}{3}A\wedge A \wedge A \right)
\fe 
suffers from a box anomaly, which can be cured by a Green--Schwarz mechanism involving BCOV/Kodaira-Spencer fields, which describe closed strings from the topological string perspective. The action for these fields takes the form 
\ie \frac{1}{2} \int_X(\bar{\partial} \mu) \partial^{-1}(\mu \vee \Omega)+\frac{1}{6} \int_X \mu \vee \mu \vee \mu \vee \Omega\fe 
where the Beltrami differential field 
\ie \mu \in \Omega^{0,1}(X, T X)\fe 
satisfies $\textrm{Div }\mu=0$, and where $X$ is the Calabi-Yau manifold on which the theory is defined, with holomorphic 3-form $\Omega$.  In the present case, we take $X=\C^3$ and $\Omega = dv\wedge dw \wedge dz$. 
The box anomaly is canceled by a BCOV Feynman diagram, but only for the gauge group $SO(8)$ \cite{Costello:2019jsy}.

The interaction between closed and open strings is captured by an interaction of the form 
\ie 
\frac{1}{8(2 \pi \mathrm{i})^{3 / 2}} \int_{\C^3} \eta \operatorname{Tr}({A} \partial{A}),
\fe
where
\ie 
\eta=\mu \vee \Omega.
\fe

The gauge invariant observables that we shall focus on in this work are holomorphic surface operators that are B-branes from the topological string perspective. 
In \cite{Costello:2020jbh}, Costello and Paquette defined these gauge invariant surface operators in 6d holomorphic Chern--Simons theory. They take the form
\ie \label{holsurop}
\sum_{n \geq 0} \frac{1}{n!} \int_{z_1, \ldots, z_n \in \mathbb{C}} \prod_{i=1}^n\left(\int \frac{1}{k_1^{i}!k_2^{i}!} \partial_{w_1}^{k_1^i} \partial_{w_2}^{k_2^i} A_{\bar{z}}^{a_i}\left(z_i\right) J_{a_i}\left[k_1, k_2\right]\left(z_i\right)\right),
\fe 
where the currents satisfy the OPE
\ie \label{classope}
J_b\left[l_1, l_2\right](0) J_c\left[m_1, m_2\right](z) \sim \frac{1}{z} f_{b c}^a J_a\left[l_1+m_1, l_2+m_2\right].
\fe 

As explained in \cite{Costello:2020jbh}, the current-current OPE on the holomorphic surface operator moreover gains a central extension due to the backreaction of the corresponding B-brane. The latter is represented by the Kodaira-Spencer field sourced by the branes, which is the Beltrami differential 
\ie \mu_{B R}=N \frac{\bar{v}d \bar{w} - \bar{w} d\bar{v}}{\left(v\bar{v}+w \bar{w}\right)^2} \partial_z,\fe 
where $N$ is the number of branes, whose support is taken to be the $z$-plane. 
Notably, the Beltrami differential deforms $\mathbb{C}^3\backslash \C$ to $SL_2(\mathbb{C})$ \cite{Costello:2018zrm}.
The Beltrami differential couples to the Chern--Simons fields via the interaction term 
\ie \label{closedint}
S_{B R}=\frac{1}{2} \int_{\mathbb{C}^3} A^a \mu_{B R} A_a d z \wedge d v  \wedge d w=-\frac{1}{2} N \int_{\mathbb{C}^3} \frac{\bar{v}d \bar{w} - \bar{w} d\bar{v}}{\left(v\bar{v}+w \bar{w}\right)^2}  A^a \partial_z A_a.
\fe

\subsection{Surface Defect Chiral Algebra from Anomaly Cancellation }

\begin{figure}[htbp]
\begin{center}
\begin{tikzpicture}
\begin{scope}[ shift={(0,0)}, scale=(0.7)]
\draw[solid](0,2.7) to (0,-2.7);

\draw [blue](0,-2.2) arc (-90:90:2.2);
\draw [blue] (40:2.2) to (40:3.2);
\draw [blue] (-40:2.2) to (-40:3.2);

\node at (2.5,2.3) {$a$};
\node at (2.5,-2.3) {$b$};

\end{scope}

\end{tikzpicture}

\caption{\small{The nontrivial Feynman diagram contributing to the deformation of \eqref{classope}.}}
\label{figure_anomaly2}
\end{center}
\end{figure}
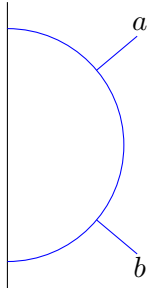

At the quantum level the  OPE \eqref{classope} required by classical gauge invariance receives a quantum correction in order to ensure that the gauge invariance of the holomorphic surface operator at the quantum level, as shown in \cite{Costello:2020jbh}. The quantum correction arises from the Feynman diagram shown in Figure 1 \ref{figure_anomaly2}, and leads to the quantum corrected OPE
\ie \label{tryang}
J_a[1,0](0) J_b[0,1](z) \sim \frac{1}{z} f_{a b}^c J_c[1,1]+\hbar \frac{1}{z} K^{f e} f_{a e}^c f_{b f}^d J_c[0,0] J_d[0,0], 
\fe 
where $K$ indicates the Killing form on the gauge Lie algebra.
This is analogous to how quantizable Wilson lines associated with representations of  polynomial loop algebras in 4d Chern--Simons theory become associated with representations of the Yangian algebra in the quantum theory. Hence, \eqref{tryang} can be understood as a multi-loop analogue of the Yangian, and one of our goals in the following sections is to derive an analogue of a rational $R$-matrix for this quantum group.

\section{B-brane Intersections in 6d Holomorphic Chern--Simons Theory}
We shall first be concerned with intersecting surface operators in 6d holomorphic Chern--Simons theory with gauge group $SO(8)$.

To compute the interaction between two holomorphic surface operators, we need to derive  the propagator of 6d holomorphic Chern--Simons theory.
We shall perform this derivation in a manner similar to the computation of the propagator of 4d Chern--Simons theory  using  the analogue of the Lorentz  gauge \cite{Costello:2017dso}, which for 6d Chern--Simons theory, is
\ie 
\partial_v A_{\vb}+\partial_w A_{\wb}+\partial_z A_{\zb}=0.
\fe 

To this end, we need to specify a boundary condition for the gauge field at infinity on the $\C$-plane  parametrized by $z$, that we shall regard as an affine patch of $\mathbb{CP}^1$. We demand that the gauge field satisfies the Dirichlet boundary condition 
\ie \label{dbc}
A|_{z=\infty}=0.
\fe 
This choice of boundary condition is analogous to 
the boundary condition used to derive rational $R$-matrices from 4d Chern--Simons theory. We shall moreover demand that gauge transformations are trivial at $z=\infty$, that is, the gauge transformation parameter satisfies $g=1$ at these boundaries. 
We also demand that the gauge field decays sufficiently fast at infinity on the $v$- and $w$-planes to justify integration by parts in these directions. 

It can be shown that these boundary conditions imply that the only possible classical solution is the trivial one, $A=0$.
To understand this, observe that the coupled equation of motion for the open-string sector in the presence of a fixed BCOV background $\mu$ is $\bar{\partial}_\mu A+A \wedge A=0$, i.e., flatness of $A$ with respect to the deformed Dolbeault operator $\bar{\partial}_\mu=\bar{\partial}+\mu^i \partial_i$ specified by the BCOV field. This equation of motion follows from varying $A$ in the coupled action described in Section \ref{2.1} and using $\textrm{Div }\mu=0$.  The solution of this equation of motion is of the form $A=-\left(\bar{\partial}_\mu g\right) g^{-1}$, and a ($\mu$-dependent) gauge transformation then sets $A$ to zero everywhere away from surface operators. 
The state space of the open-string sector of the theory is thus 1-dimensional away from surface operators. This can be understood to define a fiber functor in analogy to the case of 4d Chern--Simons theory \cite{Costello:2013zra}, which allows us to relate categorical structures arising from defects to structures associated with quantum groups, such as coproducts.  

This boundary condition is also compatible with the Green--Schwarz anomaly cancellation. The 3-form $dv\wedge dw \wedge dz$ takes the form $\frac{dv\wedge dw \wedge dz'}{z'^2}$, where $z'=-\frac{1}{z}$, in the vicinity of the boundary at $z=\infty$. The derivation in Appendix C of \cite{Costello:2021bah} then shows that the boundary condition does not introduce
new anomaly or counterterm classes. The proof compares the complex of local
Lagrangians compatible with the boundary condition to the corresponding bulk
complex after removing the divisor at $z=\infty$. Using the \(z\)-rotation grading, one finds that
in the charge sectors relevant for anomalies and counterterms the cohomology is the
same as in the bulk theory. Hence, the only possible anomalies are the usual bulk
Green--Schwarz anomaly classes, and no extra boundary anomalies appear.

 Having specified and justified the choice of boundary condition, let us proceed to compute the propagator. 
We shall express the propagator as a two-form on two copies of $\C\times \C \times \C $. We also make use of the following propagator two-form with adjoint indices removed :
\ie 
P^{a b}(v, \bar{v}, w, \bar{w}, z, \bar{z})=\delta^{a b} P(v, \bar{v}, w, \bar{w},z, \bar{z}).
\fe
The defining equations of the  propagator 
two-form are 
\begin{align}
\frac{i}{8\pi} dv \wedge dw \wedge{d} z \wedge  {d} P(v, \bar{v},w, \bar{w}, z, \bar{z}) & =\delta_{v,\bar{v},w,\bar{w}, z, \bar{z}=0} \label{propeq1} \\
\left(\partial_v \iota_{\bar{v}}+ \partial_w \iota_{\bar{w}}+ \partial_z \iota_{\bar{z}}\right) P(v,\bar{v}, w, \bar{w},z, \bar{z}) & =0 ,\label{propeq2}
\end{align}
where $\delta_{v, \bar{v},w, \bar{w}, z, \bar{z}=0} $ denotes a delta-function distribution 6-form. The second defining equation imposes the analogue of the Lorentz gauge at the level of the propagator.  
We claim that, for
$v=v'-v''$, $w=w'-w''$, and $z=z'-z''$, the propagator two-form is  
\begin{equation}
    \begin{aligned}\label{eq.propagator}
P^{ab}(v,\bar{v},z, \bar{z},w, 
\bar{w})&:=
\frac{1}{2}  \langle A^a_i (v',\bar{v}', z', \bar{z}', w', \bar{w}') A^b_j(v'',\bar{v}'', z'', \bar{z}'', w'', \bar{w}'')\rangle d x^i \wedge d x^j 
\\
&= \delta^{ab}\frac{2}{ \pi^2}\left( \vb d 
\wb \wedge d \zb +  \wb d \zb \wedge d \vb + \bar{z}
d \vb \wedge d \wb \right)\frac{1}{(v \bar{v}+w \bar{w}+z \bar{z})^{3}} 
 \;.
\end{aligned}
\end{equation}

Let us verify that this is indeed the correct propagator, verifying its normalization along the way. Firstly, the propagator goes to zero at $v,w,z=\infty$, in accordance with the boundary conditions.

Also, away from the origin, the RHS of \eqref{propeq1} ought to be zero. We can check that $dv \wedge dw \wedge dz \wedge dP $ is indeed equal to
 \ie 
dv \wedge dw\wedge dz \wedge \bigg[& \frac{2}{\pi^2} \frac{3 d\vb \wedge d\bar{w} \wedge d\bar{z}}{(|v|^2+|w|^2+|z|^2)^{3}}\\&-3 \frac{2}{\pi^2} \frac{( v d \bar{v}+w d \bar{w}+ z d \bar{z})}{(|v|^2+|w|^2+|z|^2)^{4}} \wedge(\vb d 
\wb \wedge d \zb +  \wb d \zb \wedge d \vb +\bar{z}
d \vb \wedge d \wb)\bigg]\\=0
\fe 
away from the origin. 
Moreover, we can verify \eqref{propeq2}, i.e., that 
$\left(\partial_v \iota_{{\bar{v}}}+ \partial_w \iota_{{\bar{w}}}+ \partial_z \iota_{\bar{z}}\right) P(t, w,\bar{w}, z, \bar{z})$ is proportional to 
\ie 
&\partial_v \left(\frac{-\bar{w} d\bar{z} + \bar{z}d\bar{w} }{(|v|^2+|w|^2+|z|^2)^{3}} \right) + \partial_{{z}}\left(\frac{\bar{v} d\bar{z} - \zb  d \vb  }{(|v|^2+|w|^2+|z|^2)^{3}} \right) +\partial_{{w}}\left(\frac{ -\vb  d\bar{w} + \bar{w} d\vb  }{(|v|^2+|w|^2+|z|^2)^{3}} \right) 
\\&=0.
\fe 
The normalization of the propagator is fixed by checking \eqref{propeq1} at the origin. The first step in this direction is the observation that the propagator two-form restricted to the five-sphere of radius $\epsilon$ takes the form 
\ie \label{theeq}
P(v,\vb, w, \bar{w}, z, \bar{z})=\frac{2}{ \pi^2\epsilon^6}(\vb d 
\wb \wedge d \zb +  \wb d \zb \wedge d \vb + \bar{z}
d \vb  \wedge d \wb).
\fe
If we were to integrate the propagator over a six-ball of radius $\epsilon$ (which has volume $(\pi^3\epsilon^6)/6$), then Stokes' theorem tells us that we can use \eqref{theeq} in the computation, i.e.,
\ie 
\begin{aligned}
\frac{i}{8 \pi} \int_{v \bar{v}+w \bar{w}+z \bar{z} \leq \epsilon} d v \wedge d w \wedge dz\wedge \mathrm{d} P(v,\bar{v},, w,\bar{w}, z, \bar{z}) & =\frac{i}{8 \pi} \frac{2}{ \pi^2 \epsilon^6} \int_{v \bar{v}+w \bar{w}+z \bar{z}\leq \epsilon} (-3) dv d \bar{v} dw d \bar{w} d z d \bar{z} \\
& =-\frac{i}{8 \pi} \frac{2}{ \pi^2 \epsilon^6} \int_{v\bar{v}+w \bar{w}+z \bar{z} \leq \epsilon} 3(-i2)^3d^6x \\
& =-\frac{i}{8 \pi} \frac{2}{ \pi^2} 3(-i2)^3 \frac{\pi^3}{6}=1, 
\end{aligned}
\fe
where $v=x^1+ix^2$, $w=x^3 + i x^4$, $z=x^5 + i x^6$ and $d^6x =dx^1 dx^2 dx^3 dx^4 dx^5 dx^6$. Here, a factor of $(-i2)^3$ is incurred from the measure on $\C^3$ in going from complex coordinates to Cartesian coordinates.  This completes the verification of the form and normalization of the propagator 
\eqref{eq.propagator}.

\subsection{B-brane Intersection at Leading Order in $\hbar$}

\begin{figure}
\begin{center}

\begin{tikzpicture}[tdplot_main_coords]

  \fill[white!30] (-8,-2,0) -- (-4,-2,0) -- (-4,2,0) -- (-8,2,0) -- cycle;
  \draw[thick] (-8,-2,0) -- (-4,-2,0) -- (-4,2,0) -- (-8,2,0) -- cycle;

  \fill[gray!30] (-2,-2,0) -- (2,-2,0) -- (2,2,0) -- (-2,2,0) -- cycle;
  \draw[thick] (-2,-2,0) -- (2,-2,0) -- (2,2,0) -- (-2,2,0) -- cycle;

  \fill[gray!30] (4,-2,0) -- (8,-2,0) -- (8,2,0) -- (4,2,0) -- cycle;
  \draw[thick] (4,-2,0) -- (8,-2,0) -- (8,2,0) -- (4,2,0) -- cycle;

  \node at (-3,0,0) {\huge $\times$};
  \node at (3,0,0) {\huge $\times$};

  \node[below right] at (-4,2,0) {$z$-plane};
  \node[below right] at (2,2,0) {B-brane wrapping $w$-plane};
  \node[below right] at (8,2,0) {B-brane wrapping $v$-plane};

  \fill[red] (-6,-0.4,0) circle (1pt);
  \node[below right] at (-6,-0.4,0) {$z'$};

\fill[red] (-6.5,0.6,0) circle (1pt);
\node[below right] at (-6.5,0.6,0) {$z''$};

  \fill[red] (-0.6,-0.4,0) circle (1pt);
  \node[below right] at (-0.6,-0.4,0) {$w'$};

  \fill[red] (6.4,-0.4,0) circle (1pt);
  \node[below right] at (6.4,-0.4,0) {$v''$};

\end{tikzpicture}
\end{center}
\caption{B-brane intersection as the intersection of two holomorphic surface defects in 6d Chern--Simons theory.}
\label{secondfigurr}
\end{figure}
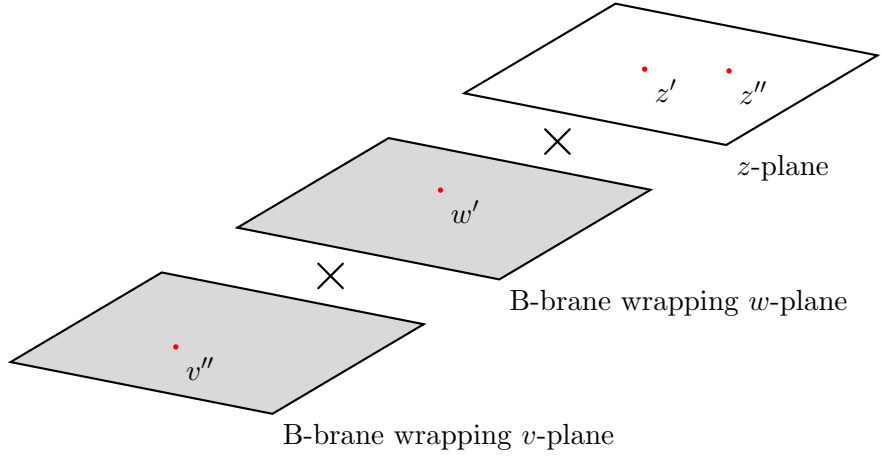

We shall consider perturbation theory around the trivial classical solution $A=0$ to compute the correlation function of intersecting surface operators, depicted in Figure \ref{secondfigurr}. We shall first consider the leading nontrivial contribution to this correlation function, which takes the form 
\ie \label{fund}
&\langle\int d^2v' \int d^2w''    A^a_{\bar{v}}(v',z',w') A^b_{\wb}(v'' ,z'' ,w'' )\rangle J_a(v')\otimes J_b(w'')\\=& \hbar \int  d^2v' \int d^2w''   \left( \frac{1}{\pi^2}\right)\bigg(\frac{2
(\bar{z}'-\bar{z}'')}{(|v'-v''|^2+|w'-w''|^2+|z'-z''|^2)^{3}}\bigg)J^a(v') \otimes J'_{a}(w'')
\fe
In order to evaluate the integrals, we first Taylor expand the currents $J^a(v')$  and $J_a(w'')$ as 
\ie 
J_a(v')=\sum_{m =0 }^{\infty} J_{am} (v')^m
\fe
\ie 
J'_a(w'')=\sum_{n =0 }^{\infty} J'_{an} (w'')^n.
\fe
The expression \eqref{fund} can then be rewritten, after the shifts $v' \rightarrow v' +v ''$ and $w'' \rightarrow w'' +w '$, as   
\ie 
 &\hbar \int d^2 v' \int d^2 w'' \left( \frac{1}{\pi^2}\right) \bigg(\frac{2
(\bar{z}'-\bar{z}'')}{(|v'|^2+|w''|^2+|z'-z''|^2)^{3}}\bigg)\sum_{m =0 }^{\infty} J^a_{m} (v'+v'')^m\otimes \sum_{n =0 }^{\infty} J'_{an} (w''+w')^n
\fe 
Performing the binomial expansions $(v'+v'')^m=\sum_{k=0}^m\binom{m}{k} (v')^{m-k} v''^k$ and $(w''+w')^n=\sum_{k'=0}^n\binom{n}{k'} (w'')^{n-k'} w'^{k'}$, and using polar coordinates on the $v'$ and $w''$-planes by setting $v'=r_{v'}e^{i\theta_{v'}}$ and  $w''=r_{w''}e^{i\theta_{w''}}$, we find that only $(v'')^m$ and
$(w')^n$ contribute to the integral, which takes the form 
\ie 
&\frac{\hbar}{\pi^2} \int_0^{\infty} r_{v'} dr_{v'}  \int_0^{2\pi}  d\theta_{v'}   \int_0^{\infty} r_{w''} dr_{w''}  \int_0^{2\pi}  d\theta_{w''}   \bigg(\frac{2
(\bar{z}'-\bar{z}'')}{(r_{v'}^2+r_{w''}^2+|z'-z''|^2)^{3}}\bigg)\\&\times \sum_{m =0 }^{\infty} J^a_{m} (v'')^m\otimes \sum_{n=0 }^{\infty}J'_{an} (w')^n.
\fe 
Performing the $r_{w''}$ and $\theta_{w''}$ integrals, we arrive at
\ie \label{rr1}
& \pi  \frac{\hbar}{\pi^2} \int_0^{\infty} r_{v'}dr_{v'} \int_0^{2\pi} d\theta_{v'} \bigg(\frac{
(\bar{z}'-\bar{z}'')}{(r_{v'}^2+|z'-z''|^2)^{2}}\bigg)J^a(v'')\otimes J'_a(w')\\=& \frac{\hbar}{z'-z''}J^a(v'')\otimes J'_a(w').
\fe

 We thus find  an order $\hbar$ contribution similar in form to that of intersecting line defects in 4d Chern--Simons theory, i.e., which has the form of  a rational classical $R$-matrix. However, here, the coefficient of $z'-z''$ is a local  operator located at the intersection of the two holomorphic surface operators.   
The same result can be obtained by working in the holomorphic gauge $A_{\bar{z}}=0$, where the propagator is proportional to $\frac{1}{z'-z''} \delta^{(2)}(v'-v'')\delta^{(2)}(w'-w'')$.

The result \eqref{rr1} is of the form anticipated by Costello in \cite{Costello:2021bah}.

\tdplotsetmaincoords{70}{120}

\subsubsection{General Surface Defects and a Yang--Baxter Equation }

We have so far considered surface defects wrapping either $\mathbb{C}_v$ or $\mathbb{C}_w$. More generally, we can consider surface defects wrapping arbitrary holomorphic curves in $\mathbb{C}_v \times \mathbb{C}_w$. For instance, given a holomorphic curve 
\ie \label{curvehol}
v=\alpha w + \beta,
\fe 
we would like to compute correlators involving surface operators on this curve. 

In other words, we would like to compute the correlation function of intersecting arbitrary holomorphic curves in $\C_v \times \C_w$. In practice, it is enough to tilt only one of the surface operators in $\C_v \times \C_w$ by an angle, because of the following reasons. 
Given that we work with 6d CS defined with respect to the holomorphic 3-form
\ie 
\Omega= \textrm{d}v\wedge \textrm{d}w\wedge \textrm{d}z, 
\fe 
we notice that the theory is invariant under $SL(3,\mathbb{C})$ transformations, which is broken to an  $SL(2,\mathbb{C})$ subgroup acting on the $v$ and $w$-planes by the boundary condition \eqref{dbc}. The 
propagator we have derived depends on a Hermitian metric, which further reduces this to $SU(2)$. Since the moduli space of each curve is $\mathbb{CP}^1$,
the total moduli space of intersecting surfaces is
\ie
\left(\mathbb{C P}^1 \times \mathbb{C P}^1\right) / S U(2) 
\fe
This leaves one angle.

We now consider the case in which the second surface defect is tilted relative to the first by this angle. 
The first defect is taken to lie along the \(v\)-plane,
\[
\Sigma_1:\qquad (v,w,z)=(v',w',z'),
\]
while the second defect is parametrised by a holomorphic coordinate \(t''\) as
\[
\Sigma_2^\theta:\qquad 
(v'',w'',z'')
=
(v_0+t''\sin\theta,\;t''\cos\theta,\;z'').
\]
For \(\theta=0\), this reduces to the unrotated configuration
\[
(v'',w'')=(v_0,t''),
\]
considered previously. We shall thus compute the correlation function between an untilted surface operator and a tilted surface operator, that is,
\[
\left\langle
\int d^2v'\,
A^a_{\bar v}(v',z',w')\,J_{av}(v')
\;
\int d^2t''\,
A^b_{\bar t''}(t'',z'')\,J'_{bt}(t'')
\right\rangle.
\]
This is performed in Appendix \ref{appena}, with the result
\[
\begin{aligned}
&
\frac{\hbar}{z'-z''}
J^a_v(v''_*)
\otimes
J'_{aw}(w'),
\end{aligned}
\]
where $v''_*=v_0+w'\tan \theta$ denotes the intersection point.\footnote{One could also perform the computation with $\tan \theta$ replaced by the complex parameter $\alpha$ from \eqref{curvehol}, and arrive at the same result. }
Note that in the untilted limit \(\theta\to0\),
\[
\tan\theta\to0,
\]
and we recover the original result given in \eqref{rr1}.

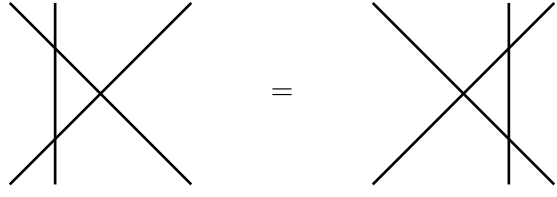
\begin{figure}[h]
\centering
\begin{tikzpicture}[line width=1pt, scale=1.2]

  \begin{scope}[xshift=0cm]
    \draw (-1,1) -- (1,-1);
    \draw (-1,-1) -- (1,1);
    \draw (-0.5,1) -- (-0.5,-1);
  \end{scope}

  \node at (2,0) {$=$};

  \begin{scope}[xshift=4cm]
    \draw (-1,1) -- (1,-1);
    \draw (-1,-1) -- (1,1);
    \draw (0.5,1) -- (0.5,-1);
  \end{scope}

\end{tikzpicture}
\caption{A Yang--Baxter equation realized in 6d Chern--Simons theory. The lines here are projections of surfaces in $\mathbb{C}^2$.}
\label{yb}
\end{figure}

Thus, we find that to leading order in $\hbar$, the result \eqref{rr1} is unchanged when the angle between intersecting surface operators is modified. This further suggests that the local operator we have derived at the intersection may indeed have the behaviour of an $R$-matrix and satisfy a Yang--Baxter equation. Indeed, one can consider the configuration of surface operators given in Figure \ref{yb}, where, at least to leading order in $\hbar$, the local operator at each surface operator intersection has the same form. Similar to the  conjecture in \cite{Costello:2021bah} for 6d CS on twistor space, a Yang--Baxter equation should hold on $\C^3$ even without the presence of diffeomorphism invariance, since one can check that  OPEs of the local operators at the intersections of defects are nonsingular since they are at different points on $\mathbb{C}_z$, which would suggest that the two configurations in Figure \ref{yb} are equivalent. Although we have found evidence consistent with a Yang--Baxter equation, a conclusive proof would require higher-order computations. In Section \eqref{coproyb}, we shall provide an indirect algebraic argument for such a Yang--Baxter equation, that requires us to first define an analogue of a coproduct for the algebra \eqref{tryang}.

\subsection{B-brane Fusion}\label{6coprod}

In 4d and 5d Chern--Simons theories, the OPEs of line and surface operators correspond to coproducts of Yangians and their affine analogues. Hence, it is reasonable to expect that the OPE of surface operators in 6d Chern--Simons theory should furnish a coproduct for the quantum algebra defined by \eqref{tryang}.

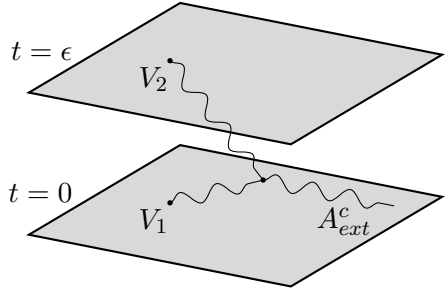
\begin{figure}
\begin{center}
\begin{tikzpicture}[tdplot_main_coords]

  \fill[gray!30] (-2,-2,-1) -- (2,-2,-1) -- (2,2,-1) -- (-2,2,-1) -- cycle;
  \draw[thick] (-2,-2,-1) -- (2,-2,-1) -- (2,2,-1) -- (-2,2,-1) -- cycle;

  \fill[gray!30] (-2,-2,1) -- (2,-2,1) -- (2,2,1) -- (-2,2,1) -- cycle;
  \draw[thick] (-2,-2,1) -- (2,-2,1) -- (2,2,1) -- (-2,2,1) -- cycle;

  \draw[decorate, decoration={snake, amplitude=1mm, segment length=5mm}] (0,-1,-1) -- (1,1,0);
  \draw[decorate, decoration={snake, amplitude=1mm, segment length=5mm}] (0,-1,1) -- (1,1,0);

  \draw[decorate, decoration={snake, amplitude=1mm, segment length=5mm}] (1,1,0) -- (1,3,0);
  \node[right] at (2.1,2.3,0) {\(A^c_{{ext}}\)};

  \fill[black] (1,1,0) circle (1pt);
 
  \fill[black] (0,-1,-1) circle (1pt);
  \node[above left=2pt] at (1.4,0,-1) {\(V_1\)};
\node[above left=2pt] at (1,-1.7,-1) {\(t=0\)};

  \fill[black] (0,-1,1) circle (1pt);
  \node[above left=2pt] at (1.4,0,1) {\(V_2\)};
\node[above left=2pt] at (1,-1.7,1) {\(t=\epsilon\)};
\end{tikzpicture}
\end{center}
\caption{B-brane fusion in 6d holomorphic Chern--Simons theory.}
\label{figureope}
\end{figure}
We shall now compute the OPE of holomorphic surface operators, as depicted in Figure \ref{figureope}. We first consider the simplest surface operators corresponding to $k_1=0$ and $k_2=0$ in \eqref{integd4}.
We would like to compute the following Feynman amplitude
\ie \label{fusionamp}
\frac{i\hbar }{8\pi}\int_{v_1}\int_{v_2}\int_{v,\bar{v},w,\bar{w},z,\bar{z}} J^a(v_1) J'^b(v_2) f_{abc}P(v_1-v,w,z)\wedge dv \wedge dw \wedge dz \wedge A^c(v,w,z) \wedge P(v_2-v,w-\epsilon,z ).
\fe
Note that the propagators $P(v_1-v,w,z)$ and $P(v_2-v,w-\epsilon,z)$ in this expression do not involve terms proportional $(\bar{v}_1-\bar{v}) d\bar{w}\wedge d\bar{z}$ and $(\bar{v}_2-\bar{v}) d(\bar{w}-\epsilon)\wedge d\bar{z}$, since we are integrating over $v_1$ and $v_2$. 

Performing the shift of variables $v_1\rightarrow v_1 + v$ and $v_2 \rightarrow v_2 + v$ gives  
\ie \label{fusionamp2}
\frac{i\hbar }{8\pi}\int_{v_1}\int_{v_2}\int_{v,\bar{v},w,\bar{w},z,\bar{z}} J^a(v_1+v) J'^b(v_2+v) f_{abc}P(v_1,w,z)\wedge dv \wedge dw \wedge dz \wedge A^c(v,w,z) \wedge P(v_2,w-\epsilon,z ).
\fe
Taylor expanding the currents as 
\ie 
J_a(v_1+v)=\sum_{m=0}^{\infty} J_{am} (v_1+v )^m
\fe
\ie 
J'_a(v_2+v)=\sum_{n=0}^{\infty} J'_{an} (v_2+v )^n. 
\fe
Performing the binomial expansions $(v_1+v)^m=\sum_{k=0}^m\binom{m}{k} v_1^{m-k} v^k$ and $(v_2+v)^n=\sum_{k'=0}^n\binom{n}{k'} v_2^{n-k'} v^{k'}$, we find that the expression  \eqref{fusionamp} is equal to
\ie \label{fusionamp2}
\frac{i\hbar }{8\pi}\int_{v_1}\int_{v_2}\int_{v,\bar{v},w,\bar{w},z,\bar{z}} J^a(v) J'^b(v) f_{abc}P(v_1,w,z)\wedge dv \wedge dw \wedge dz \wedge A^c(v,w,z) \wedge P(v_2,w-\epsilon,z ).
\fe

Now, since 
\ie 
\int r_v dr_v d\theta_v \frac{1}{(|v|^2 + |w|^2 +|z|^2)^3} = \frac{\pi}{2}\frac{1}{(|w|^2+|z|^2)^2},
\fe 
we have 
\ie \label{fusionamp2b}
\frac{i\hbar }{8\pi}\frac{\pi^2}{4}\int_{v,\bar{v},w,\bar{w},z,\bar{z}} J^a(v) J'^b(v) f_{abc}P'(w,z)\wedge dv \wedge dw \wedge dz \wedge A^c(v,w,z) \wedge P'(w-\epsilon,z ),
\fe
where
\ie 
P^{\prime}(w, z)=-\mathrm{d} \bar{z} \partial_w Q(w, z)+ \mathrm{~d} \bar{w} \partial_z Q(w, z) 
\fe 
with 
\ie 
Q(w,z):= -\frac{1}{\pi} \frac{1}{\left(|w|^2+|z|^2\right)} .
\fe
We can rewrite \eqref{fusionamp2b} as 
\ie\label{integd2}
 &\frac{i\hbar }{8\pi}\frac{\pi^2}{4}  f_{abc} \int_{t,w,\bar{w},z,\bar{z}}J^a(v) J'^b(v) A^c(t,w,z) \wedge d w \wedge dz \wedge d \bar{w} \wedge d \bar{z}\left(\partial_z Q(w) \partial_w Q(w-\epsilon)-\partial_w Q(w) \partial_z Q(w-\epsilon)\right)
\fe
where we have suppressed the $z$-dependence in $Q$. The expression \eqref{integd2} can be further rewritten as 
\ie\label{integd3}
 &\frac{i\hbar }{8\pi}\frac{\pi^2}{4} f_{abc} \int_{t,w,\bar{w},z,\bar{z}} J^a(v) J'^b(v) A^c(t,w,z) \wedge d w \wedge dz \wedge d \bar{w} \wedge d \bar{z} \left( \partial_z (a) + \partial_w (b)\right)
\fe
where 
\ie 
(a)=&\frac{1}{2}\left(-\partial_wQ(w) Q(w-\epsilon)+Q(w)\partial_wQ(w-\epsilon)\right) = \frac{1}{2\pi^2} \frac{-\epsilon \bar{w}^2 +|\epsilon|^2\bar{w} +\bar{\epsilon} |z|^2}{(|w|^2+|z|^2)^2(|w-\epsilon|^2+|z|^2)^2}\\
(b)=&\frac{1}{2}\left(\partial_zQ(w) Q(w-\epsilon)-Q(w)\partial_zQ(w-\epsilon)\right)  =-\frac{1}{2\pi^2}\frac{\bar{z} (|w-\epsilon|^2-|w|^2)}{(|w|^2+|z|^2)^2(|w-\epsilon|^2+|z|^2)^2}.
\fe

 The integral involving $(a)$ can be performed with the aid of the identity 
\ie \label{feyn}
\frac{1}{A^\alpha B^\beta}=\frac{\Gamma(\alpha+\beta)}{\Gamma(\alpha) \Gamma(\beta)} \int_0^1 \mathrm{~d} x \frac{x^{\alpha-1}(1-x)^{\beta-1}}{(x A+(1-x) B)^{\alpha+\beta}},
\fe
which gives us 
\ie 
\frac{-\epsilon \bar{w}^2 +|\epsilon|^2\bar{w} +\bar{\epsilon} |z|^2}{(|w|^2+|z|^2)^2(|w-\epsilon|^2+|z|^2)^2}=\frac{\Gamma(4)}{\Gamma(2)\Gamma(2)} \int_0^1 dx \frac{x(1-x)(-\epsilon \bar{w}^2 +|\epsilon|^2\bar{w} +\bar{\epsilon} |z|^2)}{(x(|w-\epsilon|^2+|z|^2)+(1-x)(|w|^2+|z|^2))^4}.
\fe
It can be shown that $x(|w-\epsilon|^2+|z|^2)+(1-x)(|w|^2+|z|^2)= |w-x\epsilon|^2 +|z|^2 + x(1-x)|\epsilon|^2$.
The integral \eqref{integd3} can then be rewritten as 
\ie \label{integd4}
 &-\frac{i\hbar }{8\pi}\frac{\pi^2}{4} J^a(v) J'^b(v)f_{abc} \int_{t,w,\bar{w},z,\bar{z}} A^c(t,w,z) \wedge d w \wedge d \bar{w}  \wedge dz \wedge d \bar{z} \partial_z\left(  a \right)
\fe
where 
\ie
(a)= \frac{1}{2\pi^2} \frac{\Gamma(4)}{\Gamma(2)\Gamma(2)} \int_0^1 dx \frac{x(1-x)(-\epsilon \bar{w}^2 +|\epsilon|^2\bar{w} +\bar{\epsilon} |z|^2)}{(|w-x\epsilon|^2 +|z|^2 + x(1-x)|\epsilon|^2)^4}.
\fe

We would now like to show that $\partial_z(a)$ is proportional to $\partial_z \delta^4(w,z)$ in the limit where $\epsilon \rightarrow 0$. This requires us to evaluate the integral of $(a)$:
\ie \label{inttr}
C:=\int dw \wedge d\bar{w} \wedge dz \wedge d\bar{z}(a)
\fe
To this end, we make the change of variables $w\rightarrow w + x\epsilon $, whereby \eqref{inttr} becomes 
\ie \label{inttr2}
\int dw \wedge d\bar{w} \wedge dz \wedge d\bar{z} \frac{1}{2\pi^2} \frac{\Gamma(4)}{\Gamma(2)\Gamma(2)} \int_0^1 dx \frac{x(1-x)(-\epsilon (\bar{w}+x\bar{\epsilon})^2 +|\epsilon|^2(\bar{w}+x\bar{\epsilon}) +\bar{\epsilon} |z|^2)}{(|w|^2 +|z|^2 + x(1-x)|\epsilon|^2)^4}
\fe 
Terms proportional to $\bar{w}^2$ and $\bar{w}$ in the numerator do not contribute to the integral, and we are left with 
\ie \label{inttr3}
\int dw \wedge d\bar{w} \wedge dz \wedge d\bar{z} \frac{1}{2\pi^2} \frac{\Gamma(4)}{\Gamma(2)\Gamma(2)} \int_0^1 dx \frac{x(1-x)( x(1-x)|\epsilon|^2\bar{\epsilon} +\bar{\epsilon} |z|^2)}{(|w|^2 +|z|^2 + x(1-x)|\epsilon|^2)^4}.
\fe 
Using polar coordinates $w=r_w e^{i\theta_w }$ and $z=r_z e^{i\theta_z }$, whereby the respective measures become $\int dw d \bar{w}= -i2 \int_0^{\infty } r_w dr_w \int_0^{2\pi }d\theta_w $ and $\int dz d \bar{z}= -i2 \int_0^{\infty } r_z dr_z \int_0^{2\pi } d\theta_z $, we can then perform the integrals straightforwardly. Performing the $r_w$ and angular integrals gives us 
\ie 
-8\int_0^1 dx \int_0^{\infty} r_z dr_z \left( \frac{((x)(1-x))^2 |\epsilon|^2\overline{\epsilon}}{(r_z^2+x(1-x)|\epsilon|^2)^3} + \frac{x(1-x)\overline{\epsilon}r_z^2}{(r_z^2+x(1-x)|\epsilon|^2)^3}\right).
\fe
The integral over $r_z$ can then be evaluated to give 
\ie 
-\frac{4}{\epsilon}\int_0^1 dx =-\frac{4}{\epsilon}.
\fe
An analysis analogous to that presented above shows that the term with $(b)$ does not contribute, as it involves performing an integral of the form $\int d^2z (\bar{z} f(|z|)) $ which gives zero.

Hence, we find that the Feynman amplitude \eqref{integd4} evaluates to 
\ie \label{integd5}
 &\frac{i\hbar \pi }{ 8}  f_{abc} \int dv d\bar{v} J^a(v) \otimes J'^b(v) \partial_z A^c(t,w,\bar{w},z,\bar{z}), 
\fe
where we have restored the appropriate factor of $\hbar$. This shows that, at least when restricting the open string sector, the OPE of the surface operators is finite and well-defined, and leads to another surface operator that couples to the holomorphic derivative of the gauge field. The form of this result indicates that the coproduct structure of the usual Yangian that can be derived from 4d Chern--Simons theory is present in a chiralized form in 6d Chern--Simons theory.

In fact, the computation above can be generalized straightforwardly to the case of $k_1$ and $k_2$ greater that zero in \eqref{holsurop}, whereby one finds that the class of surface operators defined in \eqref{holsurop} is closed under OPEs. A special case for surface defects involving holomorphic derivatives along the $w$-plane is derived explicitly in Appendix \ref{surffope}. There, it is shown that the surface operator that arises from computing their OPE is 
\ie
 \hbar \sum_{m, n \geq 0}{c_{m,n}}\tilde w^{-m-n-1}\int_{v_2}\pa_{z_2}A^c_{ext} \textrm{ } (J^a[0,m](v_2)J^b[0,n](v_2) f_{abc}),
\fe
where
\ie
c_{m,n}=\frac{i}{2\pi }(-1)^{n} \frac{(m+n)!}{m!n!}.
\fe
One may compare this result with an analogous coproduct for the deformed double current algebra that was computed using OPEs of line operators  in 5d Chern--Simons theory (for semi-simple gauge Lie algebra) in \cite{Ashwinkumar:2024vys} (Section 3.1), and was derived through rigorous methods in \cite{Gaiotto:2023ynn} (equation 5.4). The fact that the coefficients of the OPEs in the 5d and 6d CS theories are the same suggests that the chiral algebra associated with surface defects in 6d holomorphic CS theory can also be thought of as a chiralization of the deformed double current algebra. A similar formula also arises in the study of OPEs of line operators in 3d topological-holomorphic field theories in \cite{Dimofte:2025oqf} (equation 5.8).

\subsubsection{Contribution of B-brane Backreaction to OPE}
We also would like to consider the contribution of the closed string fields to the OPE computation. Since our main focus is on the open string sector, we shall be brief. To deduce the effect of the closed string fields, we need to take into account the interaction term 
\eqref{closedint}. 

Let us consider one surface operator at $z=w=0$ and one more at $z=0$ and $w=\epsilon$ (both with $k_1=0$ and $k_2=0$ for simplicity), and consider the effect of the backreaction of the former. The OPE computation now involves an additional term of the form 
\ie 
-\frac{1}{2} N \int_{\C^3} \frac{\bar{w} d \bar{z}-\bar{z} d \bar{w}}{\left(|w|^2+|z|^2\right)^2} A^a(w,z,v) \partial_v A^a(w,z,v)\int_{\mathbb{C}_{v_2}\times \{z=0\} \times \{w=\epsilon\}} d^2v_2A_b (\epsilon, 0,v_2)J^b(v_2),
\fe 
where the first factor arises from the backreaction and the second factor is the coupling of the holomorphic surface operator. 

Let us consider the contraction between 
$A^a(w,z,v)$ and $A_b (\epsilon, 0,v_2)$, whereby these fields are replaced by the propagator 
\ie 
\delta^{a}_{\textrm{ }b}\frac{2}{ \pi^2}\frac{\left( (\vb-\vb_2) d 
(\wb -\bar{\epsilon}) \wedge d \zb +  (\wb-\bar{\epsilon}) d \zb \wedge d (\vb-\vb_2) + \bar{z}
d (\vb-\vb_2) \wedge d (\wb-\bar{\epsilon}) \right)}{(|v-v_2|^2+|w-\epsilon|^2+z \bar{z})^{3}}. 
\fe
We can then proceed to integrate over $v_2$.
This results in an expression that is analogous in form to \eqref{fusionamp2b}, and which we know how to compute to give a finite answer. One can likewise consider the contraction between $\partial_vA^a(w,z,v)$ and $A_b (\epsilon, 0,v_2)$, and a similar computation also gives us a finite answer. In this manner, we can deduce that the backreaction of the holomorphic surface operators interpreted as B-branes
does not affect the finiteness of their OPEs.

\subsection{The Yang--Baxter Equation and Coproduct of Surface Defect Chiral Algebra }\label{coproyb}

We now provide an alternative perspective as to why the local operator that arises at the intersection of two surface operators in 6d holomorphic Chern--Simons theory can be identified with an $R$-matrix that is the solution to a Yang--Baxter equation, by relating the latter to the coproduct for the algebra \eqref{tryang}. Analogous arguments to that which follow have been used to understand a Yang--Baxter equation that arises from 5d Chern--Simons theory in \cite{Haouzi:2024qyo}.

Firstly, we recall that, from the perspective of topological holography, the modes of the bulk ghost field at the locus of a surface operator form a BRST/Chevalley--Eilenberg algebra encoding the residual gauge symmetries of 6d Chern--Simons theory. BRST invariance implies that this algebra is related by Koszul duality to a universal algebraic structure, denoted by \(\mathcal A_{\hbar}\), which governs local operators on surface operators in the Chern--Simons theory.\footnote{The derivation of Koszul duality as a consequence of BRST invariance can be found, for example, in \cite{Oh:2020hph, Paquette:2021cij}.} The local operators on a surface operator are therefore expected to furnish a representation of \(\mathcal A_{\hbar}\), which ought to be associative in the vertex-algebraic sense.
In other words, there exists an algebra homomorphism 
\ie 
\rho : \mathcal{A}_{\hbar} \rightarrow B,
\fe
where $B$ is the chiral algebra of local operators on a surface defect. 

Since we have seen that we are able to consistently fuse surface defects, we moreover deduce that there exists an algebra homomorphism, 
\ie \Delta: \mathcal{A}_{\hbar} \rightarrow \mathcal{A}_{\hbar} \widehat{\otimes} \mathcal{A}_{\hbar},\fe
which is just the coproduct,\footnote{Here, the completed tensor product is utilized in order to ensure that the image of the map may be given by a sum of infinitely many terms. } and which endows $\mathcal{A}_{\hbar}$ with the structure of a coalgebra. 

Now, the expectation is that we can define a universal $R$-matrix, denoted $\mathcal{R}\in \mathcal{A}_{\hbar} \widehat{\otimes}  \mathcal{A}_{\hbar}$ for the universal algebra $ \mathcal{A}_{\hbar}$. This universal $R$-matrix would satisfy the Yang--Baxter equation \ie \mathcal{R}_{12} \mathcal{R}_{13} \mathcal{R}_{23}=\mathcal{R}_{23} \mathcal{R}_{13} \mathcal{R}_{12},\fe
which is equivalent to 
 \begin{align} & \mathcal{R} \Delta(g)=\Delta^{\mathrm{op}}(g) \mathcal{R}, \quad \text { for any } g \in \mathcal{A}_{\hbar}  \label{aa} \\ & (\mathrm{id} \otimes \Delta) \mathcal{R}=\mathcal{R}_{13} \mathcal{R}_{12} \label{bb} \\ & (\Delta \otimes \mathrm{id}) \mathcal{R}=\mathcal{R}_{13} \mathcal{R}_{23}, \label{cc}\end{align}
where $\Delta^{\mathrm{op}}=\sigma \circ \Delta$ is the opposite coproduct that arises from composing the exchange operator $\sigma\left(g_1 \otimes g_2\right)=g_2 \otimes g_1$ with the coproduct.

Now, relations \eqref{bb} and \eqref{cc} have the following physical interpretation. Consider a surface operator on the $v$ plane, and two surface operators on the $w$ plane. At the two intersections will be local operators that we want to identify with $R$-matrices, and making the surface operators on the $w$-plane coincident is a well-defined operation that leads to a non-singular result, as we can deduce from the OPE of surface defects. 

What remains to be interpreted physically is the relation \eqref{aa}.
This relation can be understood to arise from the quantum gauge invariance of the local operator at the intersection of two surface defects, denoted $R$ in what follows.
Firstly, the gauge variation of the local operator receives contributions from both surface defects, and these contributions have to cancel one another. 

In the classical limit $\hbar \rightarrow 0$, the only contributions are linear. 
To visualize these contributions, we can draw a small sphere surrounding the intersection point, with gauge transformations picked up wherever the sphere intersects the two surface defects.

Denote the element paired with a fixed mode of the ghost field as $g \in \mathcal{A}_{\hbar=0}$. Each surface defect, which we know supports a chiral algebra, contributes via a contour integral of the generating current of the chiral algebra along an infinitesimal circle around the intersection point. Each of the two contours, one for each surface defect, is decomposed into a linear combination of two contours encircling the origin, radially ordered together with the intersection point. The two contour integrals thus each produce two modes $\rho(g)$, acting in the inward and outward directions respectively in radial ordering. The full classical gauge-invariance condition thus reads
\ie 
R\left(\rho^1_{}(g) \otimes \mathrm{id}+\mathrm{id} \otimes \rho^2_{}(g)\right)=\left(\rho^1_{}(g) \otimes \mathrm{id}+\mathrm{id} \otimes \rho^2_{}(g)\right) R,
\fe 
for any $g \in \mathcal{A}_{\hbar=0}$, where the $1$ and $2$ in superscripts are used to denote the different surface defects. 

In the quantum theory, the gauge-invariance condition receives nonlinear quantum corrections from nontrivial Feynman amplitudes, which we postulate to uplift the gauge-invariance condition to the relation
\ie \label{aa2}
R\Delta_{1,2} (g) = \Delta_{1,2} ^{\textrm{op}}(g) R,
\fe
for any $g\in \mathcal{A}_{\hbar}$, where 
\ie 
\Delta_{1,2}:= (\rho_1\otimes \rho_2)\Delta \textrm{   }:\textrm{   } \mathcal{A}_{\hbar} \rightarrow B \widehat{\otimes} B. 
\fe
The relation \eqref{aa2} is  a manifestation of \eqref{aa}.

In fact, in 5d Chern--Simons theory an analogous ``mixed" coproduct arises from demanding quantum gauge invariance at the intersection of a line and surface operator. The relevant quantum effects that contribute can be evaluated via Feynman diagram computations \cite{Oh:2021wes, Ashwinkumar:2024vys}. For 6d Chern--Simons theory, an analogous Feynman diagram computation can be performed, and this is done in Appendix \ref{interqua}, with the resulting coupling taking the form 
\begin{equation}\label{mix6}
\sum_{m\geq 0} c  f_{abc}J^a_{-m-1}[0,0] J_{-2}^b[0,m] \partial_{z_2} A_{ext}^c
\end{equation}
where 
\ie 
c=-\frac{i\hbar}{2}.
\fe
This is similar in form to the mixed coproduct derived in \cite{Ashwinkumar:2024vys, Gaiotto:2023ynn}, where the coefficient is also independent of the index, $m$. Since the compatibility of coproducts derived from 5d Chern--Simons theory arises from homomorphisms between the algebras of local operators on line and surface operators respectively \cite{Gaiotto:2023ynn}, we thus conjecture that the compatibility of \eqref{mix6} with the coproduct derived in Section \ref{6coprod} from 6d CS must be due to a vertex-algebraic \textit{endomorphism} of the underlying algebra of local operators on a surface defect. 

The proof of such an endomorphism, and more generally a conclusive proof of the relation \eqref{aa2}, would mean that the local operator at the intersection of two surface operators in 6d holomorphic Chern--Simons theory can indeed be interpreted as an analogue of an $R$-matrix that satisfies a Yang--Baxter-type equation.

\section{Surface Operator Intersections in 6d Holomorphic BF Theory}

In this section, we shall derive  local operators occurring at the intersection of surface operators in 6d holomorphic BF theory on $\C^3$. This theory has an action of the form 
\ie
\frac{i}{8\pi }\int_{\C^3} dv\wedge dw \wedge dz \wedge \textrm{Tr} \left(B\wedge F\right),
\fe
where $B$ is a one-form, and $F$ is the curvature of a one-form gauge field $A$. To ensure the cancellation of the gauge anomaly, we ought to introduce BCOV fields.  When defined on twistor space $\mathbb{PT}$, anomaly-cancellation restricts the gauge group to be $SU(2)$, $SU(3)$, $SO(8)$ or one of the exceptional groups.
The full action is 
\ie \begin{aligned} \int_{\mathbb{PT}} \operatorname{Tr}({B} F({A})) & +c \int_{\mathbb{PT}} \eta \operatorname{Tr}({A} \partial {A}) +\int_{\mathbb{PT}} \bar{\partial} \eta \partial^{-1} \eta, 
\label{bfcoup}\end{aligned}
\fe 
where $c$ is a constant computed in \cite{Costello:2021bah}.
6d holomorphic BF theory on twistor space is particularly important, as it is known to give rise to the self-dual sector of 4d Yang--Mills theory. Although we shall mostly work with 6d BF theory on $\C^3$ in what follows, we shall not show the explicit anomaly cancellation in this case, since our main goal is the local form of the correlator of intersecting surface operators for 6d BF on twistor space. 

As we shall see, the analysis of surface operators in this 6d BF theory is  similar to the case of 6d holomorphic Chern--Simons theory, at least on $\mathbb{C}^3$.
This is not a coincidence, since  6d holomorphic BF theory for the group $G$ can be understood as 6d holomorphic Chern--Simons theory for the tangent bundle group $TG$,\footnote{The Green--Schwarz anomaly cancellation for holomorphic BF theory on $\C^3$ may be derivable by generalizing the results of \cite{Costello:2019jsy} to the case of the tangent bundle group $TG$.} whose properties we shall recall now.  This is the semi-direct product group 
\ie T G \simeq G \ltimes \mathfrak{g} \equiv G \times_{\mathrm{Ad}} \mathfrak{g}\fe 
with $G$ acting on $\mathfrak{g}$, regarded as the Abelian group of translations on itself, via the adjoint representation. The corresponding Lie algebra is the ``semi-direct" sum 
\ie 
\mathfrak{t g} \equiv \mathfrak{g} \oplus_{\mathrm{ad}} \mathfrak{g}_{A b},
\fe
where $\oplus_{\mathrm{ad}}$ means that the first entry acts on the second via the adjoint representation, and $\mathfrak{g}_{Ab}$ is the vector space $\mathfrak{g}$ regarded as an abelian Lie algebra with trivial commutator. The corresponding commutation relations are 
\ie \label{tgcom}
\left[j_a, j_b\right]=f_{a b}^c j_c \quad, \quad\left[j_a, p_b\right]=f_{a b}^c p_c \quad, \quad\left[p_a, p_b\right]=0.
\fe
The invariant and non-degenerate scalar product on $\mathfrak{tg}$ is defined by 
\ie \label{scalprod}
\langle j_a, p_b \rangle & =g_{ab}\\
\langle j_a, j_b \rangle & =0\\
\langle p_a, p_b \rangle & =0,
\fe 
where $g_{ab}$ denotes the invariant Killing-Cartan form on $\mathfrak{g}$ defined as $g_{ab}=\textrm{Tr}(j_a j_b)$.

Now, for 6d holomorphic BF theory, the holomorphic surface operators have the form
\ie 
\sum_{n \geq 0} \frac{1}{n!} \int_{z_1, \ldots, z_n \in \mathbb{C}} \prod_{i=1}^n\left(\int \frac{1}{k_1^{i}!k_2^{i}!} \partial_{w_1}^{k_1^i} \partial_{w_2}^{k_2^i} A_{\bar{z}}^{a_i}\left(z_i\right) J_{a_i}\left[k_1, k_2\right]\left(z_i\right)+\int \frac{1}{l_1^{i}!l_2^{i}!} \partial_{w_1}^{l_1^i} \partial_{w_2}^{l_2^i} B_{\bar{z}}^{a_i}\left(z_i\right) \tilde{J}_{a_i}\left[l_1, l_2\right]\left(z_i\right)\right).
\fe 
Gauge invariance requires the OPEs
\ie 
J_b\left[l_1, l_2\right](0) J_c\left[m_1, m_2\right](z) &\sim \frac{1}{z} f_{b c}^a J_a\left[l_1+m_1, l_2+m_2\right]\\J_b\left[l_1, l_2\right](0) \tilde{J}_c\left[m_1, m_2\right](z) &\sim \frac{1}{z} f_{b c}^a \tilde{J}_a\left[l_1+m_1, l_2+m_2\right].
\fe 
We would like to compute the correlation function for intersections of these surface operators to leading order in $\hbar$.

To this end, we shall need the propagator for 6d holomorphic BF theory. 
As in 6d holomorphic CS, we pick the Dirichlet boundary condition for the gauge field $A$, as well as the field $B$. The equation of motion for $A$ does not involve the closed string field in this case. As a result, we find that the classical solution for the gauge field is $A=0$. It then follows that the classical solution for $B$ is $B=0$. The open string background is thus trivial away from defects.  

The propagator can be computed in the analogue of the Lorentz gauge to be
\begin{equation}
\begin{aligned}\label{eq.propagator}
P^{ab}(v,\bar{v},z, \bar{z},w, 
\bar{w})&:=
\frac{1}{2}  \langle B^a_i (v',\bar{v}', z', \bar{z}', w', \bar{w}') A^b_j(v'',\bar{v}'', z'', \bar{z}'', w'', \bar{w}'')\rangle d x^i \wedge d x^j 
\\
&= \delta^{ab}\frac{2}{ \pi^2}\left( \vb d 
\wb \wedge d \zb +  \wb d \zb \wedge d \vb + \bar{z}
d \vb \wedge d \wb \right)\frac{1}{(v \bar{v}+w \bar{w}+z \bar{z})^{3}} 
 \;.
\end{aligned}
\end{equation}
The correlator of intersecting surface operators in 6d holomorphic BF theory thus has the expansion
\ie \label{bfoperat}
1+ \frac{\hbar}{z'-z''}\left(J^a(v'')\otimes \tilde{J}'_a(w') +\tilde{J}^a(v'')\otimes {J}'_a(w')\right)+O(\hbar^2).
\fe
An alternate derivation of this result can be obtained by starting with holomorphic Chern--Simons theory for $TG$, computing the local operator at a surface intersection as before, and then proceeding to use the scalar product defined in \eqref{scalprod}.
As before, we can also consider tilted surface operators in 6d BF theory, and compute the leading order contributions to triply intersecting surface operators as denoted in Figure \ref{yb}. The OPE of local operators at the 
intersections of these defects ought to be finite and well-defined, since the surface operators are at distinct points on the $z$-plane. 

The interpretation as $TG$ holomorphic Chern--Simons theory is moreover beneficial in computing the OPE of surface operators in the open string sector. Indeed, one can just replace the structure constant in \eqref{integd5} by the structure constant of $TG$ defined via \eqref{tgcom} to obtain the desired result. 

The contribution of closed strings to the OPE computation is a little more subtle. Recall that the coupling of the BF theory to the closed string sector takes the form \eqref{bfcoup}.
   Notably, the field $B$ does not couple to the closed string fields.

Let us now consider the contribution to the OPE from the closed string sector. We are interested in the OPE of surface operators labelled (1) and (2) for convenience of exposition. Since the open-closed string interaction has the same form as 6d CS case, and does not contain the field $B$, one contracts the backreaction contribution of surface operator (1) with the $B$ ``leg" of the $BAA$ interaction vertex, and one of the $A$ ``legs" of the interaction vertex couples to the $B$ field of surface operator (2). The remaining ``leg" forms the gauge field $A$ of the surface operator that results from fusing the operators (1) and (2). The result is finite, and the OPE is thus well-defined.

Although we derived the result \eqref{bfoperat} for 6d holomorphic BF theory on $\C^3$, the same result holds locally for the same BF theory defined on twistor space, $\mathbb{PT}=\textrm{Tot}(\mathcal{O}(1)\oplus \mathcal{O}(1)\rightarrow \mathbb{P}^1)$. In this case, the coordinate $z$ is a local coordinate on the twistor sphere $\mathbb{P}^1$. However, further corrections to  \eqref{bfoperat} are expected from the nontrivial twistor geometry, and we shall leave the derivation of such corrections to future work.
Moreover, the space of classical solutions is nontrivial in the twistor background, which suggests that we ought to find solutions of an analogue of the \textit{dynamical} Yang--Baxter equation, as described in \cite{Costello:2017dso}.
Given that this holomorphic BF theory on twistor space corresponds to self-dual Yang--Mills theory, this suggests that quantum integrability in the latter may arise from a Yang--Baxter equation solved by a local $R$-operator with leading behaviour of the form \eqref{bfoperat}. It would be interesting to pursue further investigations into this Yang--Baxter equation and its implications for the self-dual sector of Yang--Mills theory.

\appendix 

\section{Tilted Surface Operator Intersection}\label{appena}

In this appendix, we compute the correlation function of an untilted surface operator with a tilted surface operator, that is, 
\[
\left\langle
\int d^2v'\,
A^a_{\bar v}(v',z',w')\,J_{av}(v')
\;
\int d^2t''\,
A^b_{\bar t''}(t'',z'')\,J'_{bt}(t'')
\right\rangle.
\]

The pullback of the antiholomorphic gauge field to the tilted defect is
\[
A_{\bar t''}
=
\sin\theta\,A_{\bar v}
+
\cos\theta\,A_{\bar w},
\]
since
\[
d\bar v''=\sin\theta\,d\bar t'',
\qquad
d\bar w''=\cos\theta\,d\bar t''.
\]
Using
\[
\langle A_{\bar v}A_{\bar v}\rangle=0,
\]
only the mixed propagator contributes, giving
\[
\begin{aligned}
&
\left\langle
\int d^2v'\,
A^a_{\bar v}(v',z',w')\,J_{av}(v')
\;
\int d^2t''\,
A^b_{\bar t''}(t'',z'')\,J'_{bt}(t'')
\right\rangle
\\
&=
\hbar\cos\theta
\int d^2v'\,d^2t''
\left(\frac{1}{\pi^2}\right)
\frac{2(\bar z'-\bar z'')}
{
\left(
|v'-v_0-t''\sin\theta|^2
+
|w'-t''\cos\theta|^2
+
|z'-z''|^2
\right)^3
}
\\
&\hspace{5cm}\times
J^a_v(v')
\otimes
J'_{at}(t'').
\end{aligned}
\]

We now define relative coordinates
\[
U=v'-v_0-t''\sin\theta,
\qquad
W=t''\cos\theta-w'.
\]
Then
\[
|v'-v_0-t''\sin\theta|^2
+
|w'-t''\cos\theta|^2
=
|U|^2+|W|^2.
\]
The complex Jacobian is
\[
\frac{\partial(U,W)}{\partial(v',t'')}
=
\begin{pmatrix}
1 & -\sin\theta\\
0 & \cos\theta
\end{pmatrix},
\]
and hence
\[
\det_{\mathbb C}
\frac{\partial(U,W)}{\partial(v',t'')}
=
\cos\theta.
\]
Therefore the real measure transforms as
\[
d^2U\,d^2W
=
|\cos\theta|^2\,d^2v'\,d^2t'',
\]
or equivalently
\[
d^2v'\,d^2t''
=
\frac{1}{|\cos\theta|^2}\,
d^2U\,d^2W.
\]

Since
\[
w''=t''\cos\theta,
\qquad
dw''=\cos\theta\,dt'',
\]
the holomorphic one-form current transforms as
\[
J'_{at}\,dt''
=
J'_{aw}\,dw'',
\]
and therefore
\[
J'_{at}
=
\cos\theta\,J'_{aw}.
\]
Hence the factor of \(\cos\theta\) from the propagator contraction combines with the factor of \(\cos\theta\) from the current transformation to cancel the Jacobian factor,
\[
\cos\theta\cdot \cos\theta\cdot \frac{1}{|\cos\theta|^2}=1
\]
for real \(\theta\).

Solving for \(t''\) and \(v'\),
\[
t''=\frac{W+w'}{\cos\theta},
\]
and
\[
v'
=
U+v_0+t''\sin\theta
=
U+v_0+\tan\theta\,(W+w').
\]
The correlator therefore becomes
\[
\begin{aligned}
&
\hbar
\int d^2U\,d^2W
\left(\frac{1}{\pi^2}\right)
\frac{2(\bar z'-\bar z'')}
{
\left(
|U|^2+|W|^2+|z'-z''|^2
\right)^3
}
\\
&\hspace{3cm}\times
J^a_v\!\left(
U+v_0+\tan\theta(W+w')
\right)
\otimes
J'_{aw}(W+w').
\end{aligned}
\]

We now expand the currents holomorphically about the intersection point,
\[
v''_* = v_0+w'\tan\theta,
\qquad
w''_* = w'.
\]
Thus, we find that
\[
J^a_v\!\left(
U+v_0+\tan\theta(W+w')
\right)
=
J^a_v(v'_*)
+
\text{terms containing positive powers of }U,W,
\]
and similarly
\[
J'_{aw}(W+w')
=
J'_{aw}(w')
+
\text{terms containing positive powers of }W.
\]
Using polar coordinates on the \(U\)- and \(W\)-planes, all positive holomorphic powers integrate to zero by angular integration. Hence only the constant terms survive, yielding
\[
\begin{aligned}
&
\hbar
\int d^2U\,d^2W
\left(\frac{1}{\pi^2}\right)
\frac{2(\bar z'-\bar z'')}
{
\left(
|U|^2+|W|^2+|z'-z''|^2
\right)^3
}
J^a_v(v''_*)
\otimes
J'_{aw}(w')
\\
&=
\frac{\hbar}{z'-z''}
J^a_v(v''_*)
\otimes
J'_{aw}(w').
\end{aligned}
\]

\section{OPE of Surface Defects in 6d CS}\label{surffope}

The Feynman amplitude that we are interested in evaluating is 
\ie 
\begin{aligned}
\frac{i}{8\pi}f_{abc} \int_{V_0,V_1,V_2} & \frac{J^a[0,m](v_0)}{m!} \frac{J^b[0,n](v_1) }{n!} dv_2 d w_2 d z_2 A^c_{e x t}\\  & \times \delta^{(2)}\left(z_0\right) \delta^{(2)}\left(w_0\right) \partial_{w_0}^m P_{02}  \delta^{(2)}\left(z_1\right) \delta^{(2)}\left(w_1-\tilde{w}\right) \partial_{w_1}^n  P_{12},
\end{aligned}
\fe 
where $P_{ij}$ is defined as 
\ie \label{pij}
P_{ij}:= P(v_i-v_j,\bar{v}_i-\bar{v}_j, w_i-w_j , \bar{w}_i-\bar{w}_j ,z_i-z_j,\bar{z}_i-\bar{z}_j).
\fe 
In particular, we are interested in deriving the coupling to the holomorphic derivative $\partial_{z_2}A^c_{ext}$.
We suppress the overall factor of $\frac{i}{8\pi}$ and restore it at the end. 

We shall first consider the $V_0$ integral,
\ie 
\int_{V_0} J^a[0,m](v_0)\delta^{(2)}\left(z_0\right) \delta^{(2)}\left(w_0\right) \partial_{w_0}^m  P_{02}
\fe
which is equivalent to
\ie
(-1)^{2m} \left(\frac{1}{-i2}\right) \frac{2}{\pi^2} 3\cdot 4 \ldots (3+ (m-1))
\int_{V_0}& J^a[0,m](v_0) \D^{(2)}(z_0)\D^{(2)}(w_0)\bar w_2^m \\& \times \frac{d\bar{v}_0\wedge(\bar z_2d\bar w_2-\bar w_2d\bar z_2)}{{(|v_{02}|^2+|w_{02}|^2+|z_{02}|^2)}^{3+m}}.
\fe
Taylor expanding $J^a[0,m](v_0)$, performing the shift of variable $v_0 \rightarrow v_0+ v_2$, using the binomial expansion $(v_0+v_2)^m=\sum_{k=0}^m\binom{m}{k} v_0^{m-k} v_2^k$, and integrating over $v_0$, we find 
\ie\label{1}
J^a[0,m](v_2) \frac{\Gamma(2+m)}{\pi}\frac{\bar w_2^m(\bar z_2d\bar w_2-\bar w_2d\bar z_2)}{(|w_2|^2+|z_2|^2)^{m+2}}.
\fe
Now, let us consider the $V_1$ integral, which is of the form
\ie 
\int_{V_1} J^b[0,n](v_1)\delta^{(2)}\left(z_1\right) \delta^{(2)}\left(w_1-\tilde{w}\right) \partial_{w_1}^n  P_{12}
\fe
 It is given explicitly by
\ie 
(-1)^{n} \left(\frac{1}{-i2}\right) \frac{2}{\pi^2} 3\cdot 4 \ldots (3+ (n-1))
\int_{V_1}& J^a[0,n](v_1) \D^{(2)}(z_1)\D^{(2)}(w_1-\tilde{w})(\bar{\tilde{w}}-\bar w_2)^n \\& \times \frac{d\bar{v}_1\wedge(\bar z_2d\bar w_2+ (\bar{\tilde{w}} - \bar w_2 )d \bar z_2)}{{(|v_{12}|^2+|\tilde{w}-w_{2}|^2+|z_{2}|^2)}^{3+n}}.
\fe 
Taylor expanding $J^b[0,n](v_1)$, performing the shift of variable $v_1 \rightarrow v_1+ v_2$, using the binomial expansion $(v_1+v_2)^m=\sum_{k=0}^m\binom{m}{k} v_1^{m-k} v_2^k$, and integrating over $v_1$, we find 
\ie\label{2}
(-1)^nJ^b[0,n](v_2) \frac{\Gamma(2+n)}{\pi}\frac{(\bar{\tilde{w}}-\bar{w}_2)^n(\bar z_2d\bar w_2+ (\bar{\tilde{w}} - \bar w_2)d\bar z_2)}{(|\tilde{w}-w_2|^2+|z_2|^2)^{n+2}}.
\fe
Combining \eqref{1} and \eqref{2} with the external gauge field, we can set up the $V_2$ integral: 
\ie
(-1)^{n} \frac{\Gamma(2+m)}{\pi}\frac{\Gamma(2+n)}{\pi}&f_{abc}\int_{V_2}(dv_2 dw_2 dz_2)\frac{\bar w_2^m (\bar z_2d\bar w_2-\bar w_2d\bar z_2)}{(|w_2|^2+|z_2|^2)^{m+2}}  \\ &\times \frac{(\bar{\tilde w}-\bar w_2)^{n}(\bar z_2d\bar w_2+(\bar{\tilde w}-\bar w_2)d\bar z_2)}{(|\bar{\tilde w}-w_2|^2+|z_2|^2)^{n+2}}A^c_{ext} J^a[0,m](v_2) J^b[0,n](v_2).
\fe
We shall be interested in the contribution to the integral coming from expanding the external gauge field $A^c_{ext}$ to leading order in $z_2$:
\ie 
A^c_{ext}=\ldots+z_2\pa_{z_2}A^c_{ext}.
\fe
The integral can be expressed as 
\ie
-(-1)^{n} \frac{\Gamma(2+m)}{\pi}\frac{\Gamma(2+n)}{\pi}\int_{V_2} & dv_2 \frac{\bar w_2^m(\bar{\tilde w}-\bar w_2)^{n}\bar z_2 \bar{\tilde w}(z_2 \pa_{z_2} A^c_{ext})}{(|w_2|^2+|z_2|^2)^{m+2}(|\tilde w-w_2|^2+|z_2|^2)^{n+2}}|dw_2|^2|dz_2|^2\\&  \times J^a[0,m](v_2) J^b[0,n](v_2)f_{abc}
\fe

The Feynman integral formula \eqref{feyn} can be used to express the integral as 
\ie
&-(-1)^{n} \frac{\Gamma(4+m+n)}{(\pi)^2} \int^1_0 dx x^{m+1}(1-x)^{n+1}\\ &\times \int_{V_2}\frac{\bar w_2^m(\bar{\tilde w}-\bar w_2)^{n}\bar z_2 \bar{\tilde w}(z_2 \pa_{z_2} A^c_{ext})|dw_2|^2|dz_2|^2}{((1-x)(|w_2|^2+|z_2|^2)+x(|\tilde w-w_2|^2+|z_2|^2))^{m+n+4}} J^a[0,m](v_2) J^b[0,n](v_2)f_{abc}.
\fe
We can rewrite the denominator of this expression as 
$\left(\left|w_2-x \tilde{w}\right|^2+\left|z_2\right|^2+x(1-x)|\tilde{w}|^2\right)^{m+n+4}$. Performing the shift $w_2 \rightarrow w_2 +x \tilde{w}$, we obtain
\ie
\int^1_0 dx x^{m+1}(1-x)^{n+1}\int_{V_2}\pa_{z_2} A^c_{ext}\frac{(\bar w_2+x\bar{\tilde w})^m((1-x)\bar{\tilde w}-\bar w_2)^{n}|z_2|^2\bar{\tilde w}}{(|w_2|^2+|z_2|^2+x(1-x)|\tilde w|^2)^{m+n+4}}|dw_2|^2|dz_2|^2 \tilde{c}_{mn,c},
\fe
where 
\ie 
\tilde{c}_{mn,c} (v_2) = -(-1)^{n} \frac{\Gamma(4+m+n)}{(\pi)^2} J^a[0,m](v_2) J^b[0,n](v_2)f_{abc}. 
\fe 
The terms in the numerator proportional to powers of $\bar{w}_2$ do not contribute due to the integral over the angular component of $w_2$. Thus, we arrive at 
\ie
\bar{\tilde w}^{m+n+1}\int^1_0 dx x^{m+1}(1-x)^{n+1}\int_{V_2}\pa_{z_2} A^c_{ext}\frac{|z_2|^2}{(|w_2|^2+|z_2|^2+x(1-x)|\tilde w|^2)^{m+n+4}}|dw_2|^2|dz_2|^2 \tilde{c}_{mn,c} .
\fe
We can now use the polar coordinates $z=r_ze^{i\theta_z}$ and $w=r_we^{i\theta_w}$, and integrate over $\theta_z$ and $\theta_w$ to find 
\ie
&- \bar{\tilde w}^{m+n+1}\int^1_0dx x^{m+1}(1-x)^{n+1}\int_{v_2}\pa_{z_2}A^c_{ext}\int^\infty_0\int^\infty_0\frac{16\pi^2r^3_zr_w}{(r_z^2+r_w^2+x(1-x)|\tilde w|^2)^{m+n+4}}dr_zdr_w \tilde{c}_{mn,c}\\
=~&- \frac{4\pi^2}{\prod_{l=1}^3(l+m+n)}{\tilde w}^{-m-n-1}\int_{v_2}\pa_{z_2}A^c_{ext}\int^1_0 dx \textrm{ } \tilde{c}_{mn,c}\\
=~&- \frac{4\pi^2}{\prod_{i=1}^3(l+m+n)}{\tilde w}^{-m-n-1}\int_{v_2}\pa_{z_2} A^c_{ext} \text{ }\tilde{c}_{mn,c}.
\fe

Restoring the overall factor of $\frac{i}{8\pi}$, the surface operator resulting from fusion thus has a coupling to $\pa_{z_2} A^c_{ext}$ of the form
\ie
 \hbar \sum_{m, n \geq 0}{c_{m,n}}\tilde w^{-m-n-1}\int_{v_2}\pa_{z_2}A^c_{ext} \textrm{ } (J^a[0,m](v_2)J^b[0,n](v_2) f_{abc}),
\fe
where
\ie
c_{m,n}=\frac{i}{2\pi }(-1)^{n} \frac{(m+n)!}{m!n!}.
\fe

\section{Quantum Corrections to Gauge Invariance of Surface Defect Intersection}\label{interqua}

The integral we would like to evaluate is
\ie 
&\sum_m\frac{i\hbar}{8\pi} \left(\frac{1}{-i2}\right)^2\sum_n \sum_l J^a_{-n-1}[0,0]J^b_{-l-1}[0,m] f_{abc}\int_{V_2}dv_2\wedge dw_2 \wedge dz_2 \\ & \wedge \int_{V_0} \delta^{(2)}(v_0) \delta^{(2)}(z_0) w_0^n P_{02} dw_0  \wedge \int_{V_1} \delta^{(2)} (v_1-\epsilon) \delta^{(2)}(z_1) \delta^{(2)}(w_1) v_1^l  \frac{1}{m!}\partial^m_{w_1}P_{12}dv_1 A_{ext}^c (v_2,w_2,z_2),
\fe
where $P_{ij}$ was defined in \eqref{pij}.
Here, location of the vertex $V_1$ has been constrained to be $v_1=\epsilon$ in order regulate the UV divergence of the diagram.
We have included a factor of $(-1/i2)^2$, since the measures of the surface operators are defined as $\frac{dw d\bar{w}}{-i2}$ and $\frac{dv d\bar{v}}{-i2}$.
We shall first consider the $V_0$ integral, which is of the form
\ie 
\frac{1}{i2}\frac{2}{\pi^2}\int_{\mathbb{C}_{w_0}} d^2w_0 \frac{w_0^n (\bar{z}_2 d\bar{v}_2 - \bar{v}_2  d\bar{z}_2)}{(|v_2|^2 +|z_2|^2 +|w_{02}|^2)^{3}} 
\fe 
Making the shift of variable $w_0 \rightarrow w_0 + w_2 $, we have
\ie 
& \frac{1}{i2} \frac{2}{\pi^2}\int_{\mathbb{C}_{w_0}} d^2w_0 \frac{(w_0+w_2)^n (\bar{z}_2 d\bar{v}_2 -\bar{v}_2 d\bar{z}_2)}{(|v_2|^2 +|z_2|^2 +|w_{0}|^2)^{3}} \\=&\frac{1}{i2} \frac{2}{\pi^2}\int_{\mathbb{C}_{w_0}} d^2w_0 \frac{w_2^n (\bar{z}_2 d\bar{v}_2 -\bar{v}_2 d\bar{z}_2)}{(|v_2|^2 +|z_2|^2 +|w_{0}|^2)^{3}} ,
\fe
due to the vanishing of the integral over the angular coordinate of $w_0$ of positive powers of $w_0$. 
Utilizing the polar coordinates $w_0=r_{w_0} e^{i\theta_{w_0}}$, we find 
\ie \label{exa}
- \frac{1}{\pi} \frac{w_2^n (\bar{z}_2d\bar{v}_2 -\bar{v}_2 d\bar{z}_2)}{(|z_2|^2 +|v_2|^2)^{2}}.
\fe

The $V_1$ integral 
\ie 
\frac{i}{2}\int_{V_1}v_1^l\delta^{(2)} (v_1-\epsilon )\delta^{(2)}(z_1)\delta^{(2)}(w_1)  \partial_{w_1}^mP_{12}
\fe 
can be evaluated easily due to the presence of the three delta functions. It is equivalent to
\ie \label{exb}
\frac{2}{\pi^2}\frac{i}{2} \bar{w}_2^m (3\cdot 4 \cdot 5 \ldots (3+m-1))\frac{(\epsilon^l (-\bar{z}_2 d\bar{v}_2  d\bar{w}_2  - \bar{w}_2 d\bar{z}_2 d\bar{v}_2 +(\bar{\epsilon} - \bar{v}_2) d\bar{w}_2 d\bar{z}_2 ))}{(|\epsilon-v_2|^2+|z_2|^2+|w_2|^2)^{3+m}}.
\fe 

Our interest is in the contribution to the integral coming from expanding the external gauge field $A^c_{ext}$ to leading order in $z_2$:
\ie 
A^c_{ext}(z_2) = \ldots + z_2\partial_{z_2} A^c_{ext}.
\fe
Hence, we are restricted to the case of $n=m$. Combining the expressions \eqref{exa} and \eqref{exb} while suppressing numerical coefficients that will be restored later, the $V_2$ integral is
\ie 
\epsilon^l\int_{V_2} dv_2 dw_2 dz_2 \frac{ |w_2|^{2m} (\bar{z}_2d\bar{v}_2 -\bar{v}_2 d\bar{z}_2)}{(|z_2|^2 +|v_2|^2)^{2}}\frac{-\bar{z}_2 d\bar{v}_2  d\bar{w}_2  - \bar{w}_2 d\bar{z}_2 d\bar{v}_2 +(\bar{\epsilon} - \bar{v}_2) d\bar{w}_2 d\bar{z}_2 }{(|\epsilon-v_2|^2+|z_2|^2+|w_2|^2)^{3+m}}A^c_{ext}.
\fe
This integral can be reexpressed as 
\ie 
\int_{V_2} \frac{\left|w_2\right|^{2 m}\left|z_2\right|^2\left(\epsilon^l\right)\bar{\epsilon}}{{({|v_2|^2+\left|z_2\right|^2})}^{2} ({|\epsilon-v_2 |^2+\left|w_2\right|^2+\left|z_2\right|^2})^{3+m}} \left|d v_2\right|^2 \left|d w_2\right|^2\left|d z_2\right|^2.
\fe 
We shall consider the case where $\epsilon \in \mathbb{R}$, and moreover, we shall only be interested in amplitudes where there is no $\epsilon$-dependence; an amplitude with $\epsilon$-dependence ought to be trivial by addition of counterterms. This holds for the case of $l=1$, where rescaling all the coordinates by $\epsilon$ removes the $\epsilon$-dependence, and other choices of $l$ give rise to amplitudes that can be removed by counterterms. Equivalently, we can set $\epsilon=1$.

Now, using \eqref{feyn}, we find (suppressing coefficients involving gamma functions)
\ie 
\int_0^1 dx\int_{V_2} \frac{\left|w_2\right|^{2 m}\left|z_2\right|^2 x^{2+m}(1-x)}{(x(|1-v_2|^2 +|z_2|^2 +|w_2|^2) +(1-x)(|z_2|^2 +|v_2|^2))^{5+m}} \left|d v_2\right|^2\left|d w_2\right|^2\left|d z_2\right|^2.
\fe 
This can be shown to be equivalent to 
\ie 
\int_0^1 dx\int_{V_2} \frac{\left|w_2\right|^{2 m}\left|z_2\right|^2 x^{2+m}(1-x)}{(|z_2|^2+x|w_2|^2 +|v_2-x|^2 +x(1-x) )^{5+m}} \left|d v_2\right|^2\left|d w_2\right|^2\left|d z_2\right|^2.
\fe 
Making the shift $v_2\rightarrow v_2 +x$, we find 
\ie 
\int_0^1 dx\int_{V_2} \frac{\left|w_2\right|^{2 m}\left|z_2\right|^2 x^{2+m}(1-x)}{(|z_2|^2+x|w_2|^2 +|v_2|^2 +x(1-x) )^{5+m}} \left|d v_2\right|^2\left|d w_2\right|^2\left|d z_2\right|^2.
\fe

Utilizing polar coordinates, and computing the integrals over angular variables, the remaining integral is
\ie 
(-i2)^3 8\pi^3\int_0^1 dx\int \frac{r_v r_w^{2m+1}r_z^3  (x^{2+m})(1-x)}{(r_z^2 + x r_w^2 + r_v^2 +x(1-x))^{4+m}}dr_v dr_w dr_z.
\fe 
After the rescaling $r_w \rightarrow r_w/\sqrt{x}$, the integral becomes 
\ie 
(-i2)^3 8\pi^3\int_0^1 dx\int \frac{r_v r_w^{2m+1}r_z^3  (x)(1-x)}{(r_z^2 +  r_w^2 + r_v^2 +x(1-x))^{4+m}}dr_v dr_w dr_z.
\fe 
We can then evaluate the remaining integrals to give a number. 

The explicit complete numerical coefficient that arises, including  factors that have been suppressed up to this point. is 
\ie
-8\pi\left(\frac{i\hbar}{8\pi }\frac{1}{m!} \right)3 \cdot 4  \ldots (3+m-1)\frac{\Gamma(5+m)}{\Gamma(2)\Gamma(3+m)}\frac{1}{(m+4)(m+3)(m+2)(m+1)}.
\fe 
 In particular, we are interested in the  $m$-dependence in this coefficient. 
Here $3 \cdot 4  \ldots (3+m-1)$ arises from \eqref{exb}, the ratio of gamma functions arises from using the Feynman integral identity \eqref{feyn}, $1/(m+4)(m+3)$ arises from the integral over $r_{z_2}$, and $1/(m+2)(m+1)$ arises from the integral over $r_{w_2}$.
Since 
\ie 
3 \cdot 4  \ldots (3+m-1)\frac{\Gamma(5+m)}{\Gamma(2)\Gamma(3+m)}=\frac{\Gamma(5+m)}{2}
\fe
and 
\ie 
\frac{\Gamma(5+m)}{(m+4)(m+3)(m+2)(m+1)}=m!,
\fe
which cancels the factor of $1/m!$ coming from the coupling of the gauge field to $J^b[0,m]$, we observe that the $m$-dependence is not present in the final numerical coefficient.

We thus arrive at
\begin{equation}
\sum_{m\geq 0} c  f_{abc}J^a_{-m-1}[0,0] J_{-2}^b[0,m] \partial_{z_2} A_{ext}^c
\end{equation}
where 
\ie 
c=-\frac{i\hbar}{2}.
\fe

\bibliographystyle{ytphys} 
\bibliography{6dCS}

\end{document}